\documentclass[letterpaper,times]{IONconf}
\usepackage{amsmath}
\usepackage{amssymb}
\usepackage{bm}
\usepackage{xcolor}
\usepackage{soul}
\usepackage{siunitx}
\usepackage{multirow}
\usepackage{tikz}
\usepackage{pgfplots} 
\pgfplotsset{compat=newest} 
\pgfplotsset{plot coordinates/math parser=false} 
\newlength\figureheight 
\newlength\figurewidth 
\usepackage{subcaption}
\definecolor{UniPDred}{RGB}{155,0,20}
\usepackage{makecell}
\usepackage{comment}

\usepackage[acronym,nomain,shortcuts]{glossaries}
\newacronym{acas}{ACAS}{assisted commercial authentication service}
\newacronym{aoa}{AOA}{angle-of-arrival}
\newacronym{awgn}{AWGN}{additive white Gaussian noise}

\newacronym{boc}{BOC}{binary offset carrier}
\newacronym{bgd}{BGD}{broadcast group delay}
\newacronym{bs}{BS}{base station}

\newacronym{cas}{CAS}{commercial authentication service}
\newacronym{cdf}{CDF}{cumulative distribution function}
\newacronym{ccdf}{CCDF}{complementary cumulative distribution function}
\newacronym{cdma}{CDMA}{code division multiple access}
\newacronym{chim}{CHIMERA}{chips-message robust authentication}
\newacronym{cn0}{$C/N_0$}{carrier to noise ratio}
\newacronym{cp}{CP}{cyclic prefix}
\newacronym{cs}{CS}{commercial service} 
\newacronym{crlb}{CRLB}{Cram\'er-Rao lower bound}

\newacronym{det}{DET}{detection error trade-off}
\newacronym{dop}{DOP}{dilution of precision} 
\newacronym{dvb}{DVB-T}{digital video broadcasting-terrestrial}

\newacronym{ecs}{ECS}{encrypted code sequence}
\newacronym{ecef}{ECEF}{Earth-Centered Earth-Fixed}
\newacronym{enu}{ENU}{east, north, up}
\newacronym{ekf}{EKF}{extended Kalman filter}

\newacronym{fa}{FA}{false alarm}
\newacronym{fll}{FLL}{frequency locked loop}

\newacronym{gdop}{GDOP}{geometric dilution of precision}
\newacronym{geo}{GEO}{geostationary Earth orbit}
\newacronym{glrt}{GLRT}{generalized likelihood ratio test}
\newacronym{gllr}{GLLR}{generalized log likelihood ratio}
\newacronym{ggto}{GGTO}{GPS to Galileo time offset}
\newacronym{gnss}{GNSS}{global navigation satellite systems}
\newacronym{gps}{GPS}{global positioning system}
\newacronym{gpst}{GPST}{GPS system time}
\newacronym{gsc}{GSC}{GNSS service center}
\newacronym{gst}{GST}{Galileo system time}

\newacronym{has}{HAS}{high accuracy service}

\newacronym{imu}{IMU}{inertial measurement unit}
\newacronym{icto}{ICTO}{inter-constellation time offset}
\newacronym{iot}{IoT}{Internet of things}

\newacronym{kf}{KF}{Kalman filter}

\newacronym{lan}{LAN}{local area network}
\newacronym{los}{LOS}{line-of-sight}
\newacronym{ls}{LS}{least squares}
\newacronym{leo}{LEO}{low earth orbit}
\newacronym{lte}{LTE}{long term evolution}

\newacronym{mac}{MAC}{message authentication code}
\newacronym{mcrlb}{MCRLB}{modified Cram\'er-Rao lower bound}
\newacronym{md}{MD}{missed detection}
\newacronym{mse}{MSE}{mean square error}
\newacronym{meo}{MEO}{medium Earth orbit}
\newacronym{mvu}{MVU}{minimum variance unbiased}

\newacronym{nma}{NMA}{navigation message authentication}
\newacronym{nmea}{NMEA}{national marine electronics association}
\newacronym{nmsb}{NMSB}{normalized mean square bandwith}
\newacronym{nlos}{NLOS}{non-line-of-sight}
\newacronym{ntp}{NTP}{network time protocol}

\newacronym{osnma}{OS-NMA}{open service \ac{nma}}
\newacronym{ofdm}{OFDM}{orthogonal frequency division multiplexing}

\newacronym{pdop}{PDOP}{position dilution of precision}
\newacronym{pdf}{PDF}{probability density function}
\newacronym{pnt}{PNT}{position navigation and timing}
\newacronym{ppm}{ppm}{parts per million}
\newacronym{prn}{PRN}{pseudo-random noise}
\newacronym{prb}{PRB}{physical resource block}
\newacronym{psd}{PSD}{power spectral density}
\newacronym{ptp}{PTP}{precision time protocol}
\newacronym{psk}{PSK}{phase shift keying}
\newacronym{pvt}{PVT}{position, velocity, and time}
\newacronym{prs}{PRS}{positioning reference signal}
\newacronym{pss}{PSS}{primary synchronization sequence}

\newacronym{qam}{QAM}{quadrature amplitude modulation}

\newacronym{re}{RE}{resource element}
\newacronym{recs}{RECS}{re-encrypted code sequence}
\newacronym{rinex}{RINEX}{receiver independent exchange format}
\newacronym{rv}{rv}{random variable}
\newacronym{rssi}{RSSI}{received signal strength indicator}

\newacronym{sca}{SCA}{spreading code authentication}
\newacronym{sce}{SCE}{spreading code encryption}
\newacronym{scer}{SCER}{secure code estimation and replay}
\newacronym{sbf}{SBF}{Septentrio binary format}
\newacronym{soop}{SOOP}{signals of opportunity}
\newacronym{sdr}{SDR}{software-defined radio}
\newacronym{sqm}{SQM}{signal quality monitoring}
\newacronym{snr}{SNR}{signal-to-noise ratio}
\newacronym{sssc}{SSSCs}{spread spectrum security codes}
\newacronym{std}{STD}{standard deviation}
\newacronym{sv}{SV}{satellites vehicle}
\newacronym{sss}{SSS}{secondary
synchronization sequence}

\newacronym{tesla}{TESLA}{timed-efficient stream loss-tolerant authentication} 

\newacronym{uere}{UERE}{user equivalent ranging error}
\newacronym{ue}{UE}{user equipment}
\newacronym{utc}{UTC}{coordinated universal time}
\newacronym{vctcxo}{VCTCXO}{voltage-controlled and temperature-controlled crystal oscillator}

\newcommand\comNL[1]{\textcolor{blue}{\slshape [NL: #1]}}

 \usepackage{graphicx}

\usepackage{url}

\usepackage{natbib}

\usepackage[hidelinks]{hyperref}

\title{Performance Limits for Signals of Opportunity-Based Navigation}

\author{ Francesco Zanirato$^1$, Francesco Ardizzon$^1$, Laura Crosara$^1$, Alessio Curzio$^2$, \\  Luca Canzian$^2$, Stefano Tomasin$^{1,3}$, and Nicola Laurenti$^{1,3}$%
    \vspace{1mm} \\
    $^1$ \textit{Department of Information Engineering, University of Padova, Padova, Italy}
    \vspace{1mm} \\
    $^2$ \textit{Qascom S.r.l, Bassano del Grappa, Vicenza, Italy}
    \vspace{1mm} \\
    $^3$ \textit{Consorzio Nazionale Interuniversitario per le Telecomunicazioni (CNIT), Parma, Italy}
    }

\begin{document}

\maketitle

\section*{biography}


\biography{Francesco Zanirato}{Francesco Zanirato received the B.Sc. degree in electronic engineering from the University of Padova in 2022. He is currently a M.Sc. student in telecommunications engineering and he is conducting his research training activity for the master thesis in the field of signals of opportunity.}

\biography{Francesco Ardizzon}{
Francesco Ardizzon (Member, IEEE) received the B.Sc. degree, the M.Sc. degree, and the Ph.D. degree in information engineering from the University of Padova, Italy, in 2016, 2019, and 2023, respectively. In 2022, he has been a Visiting Scientist with the ESA European Space Research and Technology Centre. He is currently an Assistant Professor at the University of Padova. His current research interests include authentication for global navigation satellite systems, physical layer security, and underwater acoustic communications.}

\biography{Laura Crosara}{Laura Crosara received the B.Sc. degree in information engineering and the M.Sc. degree in telecommunications engineering from the University of Padova, Italy, in 2019 and 2021 respectively, where she is currently pursuing the Ph.D. degree in Information Engineering. She is currently a Visiting Expert at the European Space Research and Technology Centre in Noordwijk, The Netherlands. Her current research interests include authentication techniques for global navigation satellite systems, physical layer security, and wireless communications.}

\biography{Alessio Curzio}{Alessio Curzio is a GNSS Receiver and Test Engineer at Qascom with one year of experience in GNSS, PNT, and signal processing. Since 2023, he has been working on the development of receivers for LEO PNT systems and using signals of opportunity. He is also involved in testing user receivers with Galileo 1st generation signals. Alessio holds a Master's degree in Telecommunications Engineering from the University of Padua (Italy), focusing his thesis on prototyping a Starlink receiver for opportunistic positioning.}

\biography{Luca Canzian}{Luca Canzian joined Qascom in 2015 and he is currently leading the R\&D domain area. He has worked in several projects with ESA, ASI, NASA, the European Commission and Industry. His main expertise involves ground-based and space-based location systems, detection and location of interference signals, navigation using signals of opportunity, inertial navigation systems, orbit determination techniques, GNSS authentication and anti-spoofing techniques. He holds a MSc degree and a PhD in Electrical Engineering from University of Padova (Italy)}

\biography{Stefano Tomasin}{Stefano Tomasin received the Ph.D. degree from the University of Padova, Italy, in 2003. He joined the University of Padova where he has been Assistant Professor (2005-2015), Associate Professor (2016-2022), and Full Professor (since 2022).  His research interests include physical layer security, security of GNSS, and signal processing for wireless communications. He is a senior member of IEEE since 2011. He is Editor of the EURASIP JWCN (since 2011) and Deputy Editor-in-Chief of the IEEE TIFS.}

\biography{Nicola Laurenti}{Nicola Laurenti is an Associate Professor at the Department of Information Engineering, University of Padua, Italy.
His research interests mainly focus on wireless network security at lower layers (physical, data link, and network), satellite navigation systems, information theoretic security, and quantum cryptography. He has led several activities related to the information security of satellite navigation systems, funded by the European Space Agency.}

\section*{Abstract}


This paper investigates the potential of non-terrestrial and terrestrial \ac{soop} for navigation applications. Non-terrestrial \ac{soop} analysis employs \ac{mcrlb} to establish a relationship between \ac{soop} characteristics and the accuracy of ranging information. This approach evaluates hybrid navigation module performance without direct signal simulation. The \ac{mcrlb} is computed for ranging accuracy, considering factors like propagation delay, frequency offset, phase offset, and \ac{aoa}, across diverse non-terrestrial \ac{soop} candidates. Additionally, Geometric Dilution of Precision (GDOP) and \ac{leo} SOOP availability are assessed. Validation involves comparing \ac{mcrlb} predictions with actual ranging measurements obtained in a realistic simulated scenario. Furthermore, a qualitative evaluation examines terrestrial \ac{soop}, considering signal availability, accuracy attainability, and infrastructure demands.

\glsresetall

\section{Introduction}
\Ac{soop} are signals that, despite being designed for communications, are opportunistically used by the receiver for navigation purposes as well.  

\ac{soop} signals encompass both satellite and terrestrial sources. Among terrestrial sources are cellular, i.e., LTE and 5G/6G networks, Wi-Fi, and Bluetooth, while satellite \ac{soop} include \ac{leo} satellite constellations like Starlink and Orbcomm.
Indeed, \Ac{soop} may be exploited to extract additional ranging estimations when \ac{gnss} signals are compromised or unavailable. In particular, terrestrial \ac{soop} offer dense deployment and bandwidth advantages, particularly beneficial in indoor settings and urban areas, where the number of visible \ac{gnss} satellites is limited. 
Moreover, \ac{leo} satellite constellations are of interest concerning the satellite \ac{soop} as they provide a good decorrelation of measurement errors, potentially assisting \ac{gnss} when a hybridized \ac{pvt} solution is computed. Also due to their limited cost with respect to \ac{meo} \ac{gnss} satellites, thousands of \ac{leo} satellites orbit around the Earth, and each of these may represent a source of ranging. 

Still, employing \ac{soop} for positioning presents challenges, as these signals are not designed for navigation but serve various purposes. For instance, retrieving pseudoranges requires information such as exact transmission time and transmitter location, which may not be trivial to obtain. For instance, while knowledge of the location of stationary transmitters is typically easy to retrieve and often already known to the receiver (e.g., the base station location in 5G/6G) when transmitters are moving, such information is often inaccurate and hard to retrieve.
Moreover, most communication systems used as \ac{soop} are not time-synchronized to a nanosecond accuracy (as \ac{gnss} are) and a time correction may not be available to the receiver. Other relevant issues are \ac{nlos} and multipath effects.

Various measurements can be drawn from \ac{soop}, such as carrier phase, time-of-arrival, \ac{aoa}, or Doppler frequency. These measurements are later processed to extract ranging information such as the pseudorange or pseudorange rate, and may be integrated with \ac{gnss}-ranging information in a hybrid navigation module. 
Indeed mixing signals from diverse systems/networks, even if not originally intended for positioning, may represent a huge opportunity for navigation. 

For instance, the use of terrestrial \ac{soop} from 5G/6G cellular or Wi-Fi networks may aid navigation in urban canyon environments. Moreover, \ac{soop} represent a valid resource in contexts where the signal navigation cannot be trusted, e.g., after the detection of jamming or spoofing attacks.
Thus, approaches relying on hybrid navigation solutions enhance \ac{pvt} resilience, availability, and accuracy.



This paper explores the usage of non-terrestrial and terrestrial \ac{soop} for navigation purposes. For non-terrestrial \ac{soop}, we employ the \ac{mcrlb} analysis to establish a connection between the characteristics of \ac{soop} and the accuracy of the ranging information estimated from each signal. This approach allows us to assess the performance potential of a hybrid navigation module without direct signal simulation. 
Specifically, we calculate the \ac{mcrlb} for ranging accuracy, considering various measurements such as propagation delay, frequency offset, phase offset, and \ac{aoa}, across different non-terrestrial \ac{soop} candidates. 
We then evaluate \ac{gdop} and the availability of various \ac{leo} \ac{soop}.
Finally, to validate the derived bounds rigorously, we compare the \ac{mcrlb} with actual ranging measurements obtained in a realistic simulated scenario.

Lastly, we provide a qualitative assessment of the positioning performance achievable by terrestrial \ac{soop}, taking into account signal availability, achievable accuracy, and infrastructure requirements.

\section{Methodology}

To identify the best \ac{soop}, either terrestrial or non-terrestrial, we consider two criteria: the positioning accuracy and the \ac{soop} availability. 

We analyze the achievable positioning accuracy of the \ac{pvt} when using different systems. This can be characterized by considering 
\begin{equation}
    \sigma_\mathrm{pos}^2 \triangleq \mathbb{E}\left[ \left(\bm{p} - \hat{\bm{p}}\right)^2 \right] \,,
\end{equation}
where $\bm{p}$ and $\hat{\bm{p}}$ are respectively the true and measured \ac{pvt} results.

Moreover, as in \cite{Reid2016LeveragingCB}, positioning accuracy can be approximated as 
\begin{equation}\label{eq:accuracy_GDOPUERE}
    \sigma_\mathrm{pos} = \sigma_\mathrm{UERE} \, \mathrm{GDOP}\,,
\end{equation}
thus it is related to both the ranging accuracy, expressed by $\sigma_\mathrm{UERE}$, that is the \ac{uere} variance, and the \ac{gdop}.
The former term pertains to the estimation accuracy of the distance between the transmitter and receiver, considering both signal processing aspects and signal characteristics. Conversely, the latter term, $\mathrm{GDOP}$, addresses the accuracy influenced by the spatial arrangement of satellites relative to the receiver, e.g., the number of visible satellites and their distribution.

Thus, given \eqref{eq:accuracy_GDOPUERE}, the best \ac{soop} will be ones minimizing $\sigma_\mathrm{UERE}$ and $\mathrm{GDOP}$. 


\section{Limits on the Accuracy of the Estimation of Non/Terrestrial SOOP Observables}

We consider a receiver acquiring signals from different \ac{leo} systems, and from each of them, it estimates the propagation delay $\tau$, the phase offset $\theta$, and the frequency offset $\upsilon$.
The objective of this analysis is to determine the best non-terrestrial \ac{leo} candidate by considering the ranging and ranging rate accuracy via \acp{mcrlb}, considering various measurements. In other terms,  \acp{mcrlb} will provide the minimum variance of the estimator associated with each measurement and \ac{leo} system under consideration.

In detail, we will consider Iridium NEXT, OrbComm, OneWeb, and Starlink as \ac{leo} candidates.

\subsection{Channel and Signal Model}

The signal at the receiver in baseband can be decomposed into useful signal, $s(t)$, and \ac{awgn} noise $w(t)$ at the receiver, having zero mean and variance $N_0$, i.e., 
\begin{equation}
    r(t) = s(t) + w(t) \: , 
\end{equation}
where $s(t)$ is defined as 
\begin{equation}
    s(t) = e^{j(2 \pi \upsilon t + \theta)}\sum_{i=1}{N}c_{i} g_{i}(t-iT-\tau) \: .
    \label{eq:signal}
\end{equation}
In detail, $g_i$ describes the waveform of the $i$th pulse, $T$ is the symbol duration, and $c_i$ denotes the message symbols. In particular, we consider the information to be modulated in (eventually complex) amplitude (e.g., \acrshort{qam}) or in phase (\acrshort{psk}).

We remark that the phase and frequency offsets are defined with respect to the receiver local oscillator, having instantaneous phase $\Phi_{\rm L}$ and frequency $f_{\rm cL}$. Thus the frequency offset is
\begin{equation}
    \upsilon = f_\mathrm{c} - f_{\rm cL}\,,
\end{equation}
modeling, for instance, also the shift induced by the Doppler effect, while the phase shift term $\theta$ is instead
\begin{equation}
    \theta = \Phi_{\rm L} - 2\pi f_\mathrm{c} \tau \: . 
    \label{phase_shift}
\end{equation}

\subsection{Modified Cramer Rao Lower Bounds for LEO Observables}
The traditional \ac{crlb} \cite{mengali1997synchronization} provides the estimator variance when all the parameters are known.
However, we aim at deriving estimates from \ac{soop} data signals, where the modulated data is not known by the receiver. 
Thus, we resort to the \ac{mcrlb} of \cite{dandrea94modified}, where for a measurement $\lambda$ and a signal in baseband
\begin{equation}
\mathrm{MCRLB}(\lambda) \triangleq \frac{N_0/2}{\mathbb{E}_{\bm{u}}\left[  \int_0^{T_0} \left| \frac{\partial s(t, \lambda,  \textit{\textbf{u}})}{\partial \lambda} \right|^2 \, dt \right]} \: , 
\label{MCRLB_baseband}
\end{equation}
where $\mathbb{E}_{\bm{u}}$ represents the expectation taken with respect to the random components of the signal (e.g., data) and $N_0$ the noise \ac{psd}.
Indeed, 
\begin{equation}
    \mathrm{CRLB}(\lambda) \geq \mathrm{MCRLB}(\lambda)\,,
\end{equation}
so the \ac{crlb} represents a more conservative version of the lower bound on variance, and the two values coincide if the unwanted parameters are known at the receiver or no unwanted parameters exist at all.

Considering the signal pattern in \eqref{eq:signal}, we compute the \ac{mcrlb} for the variance of $\tau$, $\upsilon$ and $\theta$ estimators
\begin{equation}
\mathrm{MCRLB}(\tau) =  \frac{T^2}{T_0 \xi \frac{C}{N_0}}{8\pi^2}\,,
\label{mcrlb_tau_cndr}
\end{equation}
\begin{equation}
    \mathrm{MCRLB(\theta)} =  \frac{1}{{T_0}  \frac{C}{N_0}} 2
\label{mcrlb_theta_cndr}
\end{equation}
\begin{equation}
    \mathrm{MCRLB}(\upsilon) = \frac{3}{{T_0}^3 \frac{C}{N_0}} {2\pi^2}
\label{mcrlb_upsilon_cndr}
\end{equation}
which include the \ac{cn0} (in dB-Hz), the observation time $T_0$. $\xi$ denotes the \ac{nmsb}, defined, for a signal with \ac{psd} $G(f)$, as
\begin{equation}\label{eq:xi}
\xi = T^2  \frac{ \int_{-\infty}^{+\infty}f^2  \left| G(f) \right| ^2 df}{ \int_{-\infty}^{+\infty}\left| G(f) \right| ^2 df}\,.
\end{equation}

A fourth method could involve the estimation of the \ac{aoa}, $\beta$, of one of the tones using an array of $M$ (synchronized)  equally spaced antennas, with a spacing $d$. The maximum accuracy with which the angle of arrival can be estimated as \cite{Kay97}
\begin{equation}\label{eq:mcrlb_aoa}
\mathrm{MCRLB}(\beta) \geq \frac{12}{(2\pi)^2 M  \frac{C}{N_0} T_0 \frac{M+1}{M-1} \left( \frac{L f_\mathrm{c}}{c} \right) ^2 \mathrm{sin}^2\beta}\,,
\end{equation}
which, with respect to the previously described \acp{mcrlb} also depends on $M$, the overall array length $L = d(M-1)$, and the \ac{aoa} itself.

All the \acp{mcrlb} show that as the \ac{cn0} and signal observation time increase, the estimator accuracy increases as well. Specifically, concerning the $\tau$ estimation of \eqref{mcrlb_tau_cndr}, the lower bound also depends on the actual signal bandwidth. Thus, to compute the bounds, we need the exact signal parameters. 

An important issue to be tacked is the estimation of the \ac{cn0}. In \cite{ferre2021comparison}, the authors report for systems \ac{leo}  between 20 dB-Hz and 50 dB-Hz of \ac{cn0}.
The maximum \ac{cn0} can be estimated via the link budget formula reported in \cite{formula_link_budget} here written in its linear form
\begin{equation}\label{link_budget}
C/N_0 =\mathrm{EIRP} \frac{G}{T_{\rm temp}} L_P \frac{1}{k}\,,
\end{equation}
where $\mathrm{EIRP}$ stands for the effective isotropic radiated power, which takes into account the amount of power radiated by the transmitter antenna and its radiation pattern, $G$ is the gain of the receiver antenna, $T_{\mathrm{temp}}$ represents the system temperature, $L_{\rm P}P$ is the free-space path loss, which in turn is function of the altitude of the satellite and the carrier frequency of the signal, and $k$ is the Boltzmann constant.

In particular for the path loss we considered the minimum distance over the ground receiver position, i.e., the altitude of the satellite (e.g., for Starlink \SI{550}{\kilo\meter}).
In the following, we describe each \ac{leo} signal parameter.

\paragraph{Starlink} 
Concerning Starlink, the structure of the signal has been reverse-engineered in \cite{Humphreys23Signal}. Starlink uses an \ac{ofdm} of which we know the overall symbol duration, cyclic prefix duration, number of subcarriers as well as the number and amplitude of each channel. 
Analyses on the \ac{cn0} have been provided in \cite{Osoro21Techno, Aguilar19Tradespace, delportillo2019technical, xia2019beam, ferre2022isleo} provides an estimate of $42.6$ dB-Hz for outdoor. On the other hand, this quantity turns out to be much lower than the one we derived through \eqref{link_budget} and \cite{Osoro21Techno, Aguilar19Tradespace, delportillo2019technical, xia2019beam}.
We remark that according to \cite{Humphreys23Signal}, consecutive frames from the same satellite and the same beam may be transmitted with different (not specified) powers.
The duration of a frame is about 1.33 ms of which the last 5.33 $\mu$s of silence, however, frames are not transmitted with perfect regularity (the frequency depends on the demand for data traffic) but at least one frame is always detected about every 40 ms (1 in 30) \cite{Humphreys23Signal}.


The calculation of $\xi$ for the Starlink system was performed by considering the entire channel and assuming a trapezoidal impulse shape, so \eqref{eq:xi} becomes
\begin{equation}
    \begin{split}
    \xi = T^2\frac{\sum_{i = -N/2+1}^{N/2} \int_{-\infty}^{+\infty} f^2  |G_{i}(f)|^2 \, df}{\sum_{i = -N/2+1}^{N/2} \int_{-\infty}^{+\infty} |G_{i}(f)|^2 \, df} &= T^2 \frac{\sum_{i = -N/2+1}^{N/2} \int_{-\infty}^{+\infty} f^2  |G_{0}(f-iF)|^2 \, df}{\sum_{i = -N/2+1}^{N/2} \int_{-\infty}^{+\infty} |G_{0}(f-iF)|^2 \, df} \\ &
    = T^2 \left(-\frac{1}{(2\pi)^2}  \frac{2}{T_C}  \frac{1}{T_{\rm sym}-\frac{1}{3} T_C} + \frac{{F N}^2}{12} + \frac{F^2}{6}\right)
    \end{split}
\end{equation}
where $F$ is the width of a single sub-band and $N$ is the number of subcarriers, $T_{\rm sym}$ the OFDM symbol period, $T_{\rm C}$ the OFDM chip period. 

\paragraph{OneWeb}
Only partial information on OneWeb signals is available. In \cite{10139969} the authors claim that, as for Starlink, its spectrum is divided into 8 channels of 250 MHz placed between 10.7 GHz and 12.7 GHz. Each of them implements an OFDM structure. The data used to estimate the link budget were obtained from \cite{delportillo2019technical}, \cite{xia2019beam} and \cite{Osoro21Techno}.

\paragraph{Iridium}
According to \cite{iridium_manual}, the signal transmitted by the Iridium satellites consists of $240$ channels of $\SI{31.5}{\kilo\hertz}$ each. These use a QPSK modulation whose symbols last for \SI{40}{\micro\second} and their waveforms have a $40\%$ roll-off factor. The link budget was computed by using the data reported in \cite{iridium_link_budget}.

\paragraph{Orbcomm}
The Orbcomm signal is the one using the least amount of bandwidth: the spectrum is only $\SI{4.8}{\kilo\hertz}$ wide. The modulation is a symmetric differential PSK whose pulse shape has a $40\%$ roll-off. The data for the link budget were obtained from \cite{orbcomm_manuale}.

Finally, the parameters for the \ac{leo} systems under consideration have been summarized in Table \ref{tab:signal_parameters}.

\begin{table}
    \centering
    \caption{Summary of the LEO \ac{soop} parameters derived from the literature, where N/A stands for {\em not currently available}.}
    {\footnotesize    
    \label{tab:signal_parameters}
\begin{tabular}{|c|c|c|c|c|c|}
\hline

\textbf{} & \makecell{ \textbf{Starlink}} & \textbf{OneWeb} & \textbf{Iridium} & \textbf{Orbcomm} \\ \hline

\textbf{{Modulation}} & \makecell{\\ OFDM \\ \\ $N = 1024$ \\ $T_{\text{sym}} = 4.4 \, \mu\text{s}$ \\ $T_C = 4.167 \,\text{ns}$ \\$F = 234375$ Hz\\ \\} & \makecell{\\ OFDM \\ \\ $N = $ N/A \\ $T_{\text{sym}} = $ N/A \\ $T_C = $ N/A \\$F = $ N/A \\ \\} & \makecell{QPSK \\ \\40\% roll-off \\ $T_{\text{sym}} = 40 \mu\text{s}$} & \makecell{SD-QPSK \\ \\ 40\% roll-off \\ $T_{\text{sym}} = 208.33 \mu\text{s}$} \\ \hline

\makecell{\\ {\textbf{Bandwidth}}\\ \\} & \makecell{\\240 MHz \\(8 channels)\\ \\}& \makecell{\\250 MHz \\(8 channels)\\ \\} & \makecell{\\31.5 kHz \\ (240 channels)\\ \\} & \makecell{4.8 kHz}  \\ \hline

\makecell{{\textbf{Carrier frequency}}} & \makecell{\\11.57 GHz \\ ($4$th channel) \\ \\} & 11.075 GHz & \makecell{1.621 GHz\\($120$th channel)} & 137.5 MHz \\ \hline

\makecell{ \\ {\textbf{Altitude and}} \\ \textbf{{relative p.l.}}\\ \\} & 550 km;  168.5 dB& 1200 km; 174.9 dB & 780 km; 154.5 dB & 750 km; 132.7 dB \\ \hline

\makecell{\\{\textbf{Estimation of}} \\ {\textbf{the maximum}} \\ {\textbf{achievable}} \\ $C/N_0$ \\ \\} & \makecell{109.3 dB-Hz \\  } & \makecell{105.53 dB-Hz \\   } & \makecell{80.6 dB-Hz \\  } &  \makecell{79.6 dB-Hz \\     }\\ \hline

\makecell{\\ {\textbf{Value of}} $C/N_0$ \\ {\textbf{found in the literature}}\\ \\} & 42.6 dB-Hz  & 31.9 dB-Hz  & \makecell{N/A} & \makecell{ N/A } \\ \hline

\makecell{\\ {\textbf{Beacon Length}}\\ \\} & \makecell{\\1.33 ms \\(max $D = 99.7 \%$) \\ \\} & 10 ms & \makecell{90 ms \\ (max $D = 36.8 \%$ )}& \makecell{\\ 1 s \\ (max $D = 50$ \% \\ commonly $D$ varies \\ from $6$ \% to $10$ \%)\\ \\} \\ \hline

\makecell{\\ {\textbf{Notes}}\\ \\} & \makecell{\\Presence of unmodulated \\carriers around \\ 11.325 GHz spaced \\ by 44 kHz. C/N0 varies \\ between 24 dB-Hz \\ and 36 dB-Hz  \\ \\} & \makecell{\\Unknown inner structure \\ of the OFDM. The  computation \\ of  $\mathrm{MCRLB}(\tau)$ is estimated via \\ flat spectrum model.\\ \\ } & \makecell{$-$} & \makecell{$-$}\\ \hline

\end{tabular}
}

\end{table}




\subsection{Numerical Results: MCRLBs}
The following results have been converted from the time domain to the space domain by considering  $ c = 3 \times {10^8} \si{\meter/\second}$.
In order to evaluate the dependency of ranging accuracy on both \ac{cn0} and observation time, we first consider the Starlink signal. Next, we will report the results of the comparative analysis.
Finally, to validate the use of the \acp{mcrlb} we report the results of a simulation of an acquisition from Starlink, comparing the statistic of the estimated measurements with the \ac{mcrlb}.

\begin{figure}[ht]
\centering
\subfloat[Propagation Delay, $\tau$. \label{fig:mcrb_tau}]{\includegraphics[width=0.5\columnwidth]{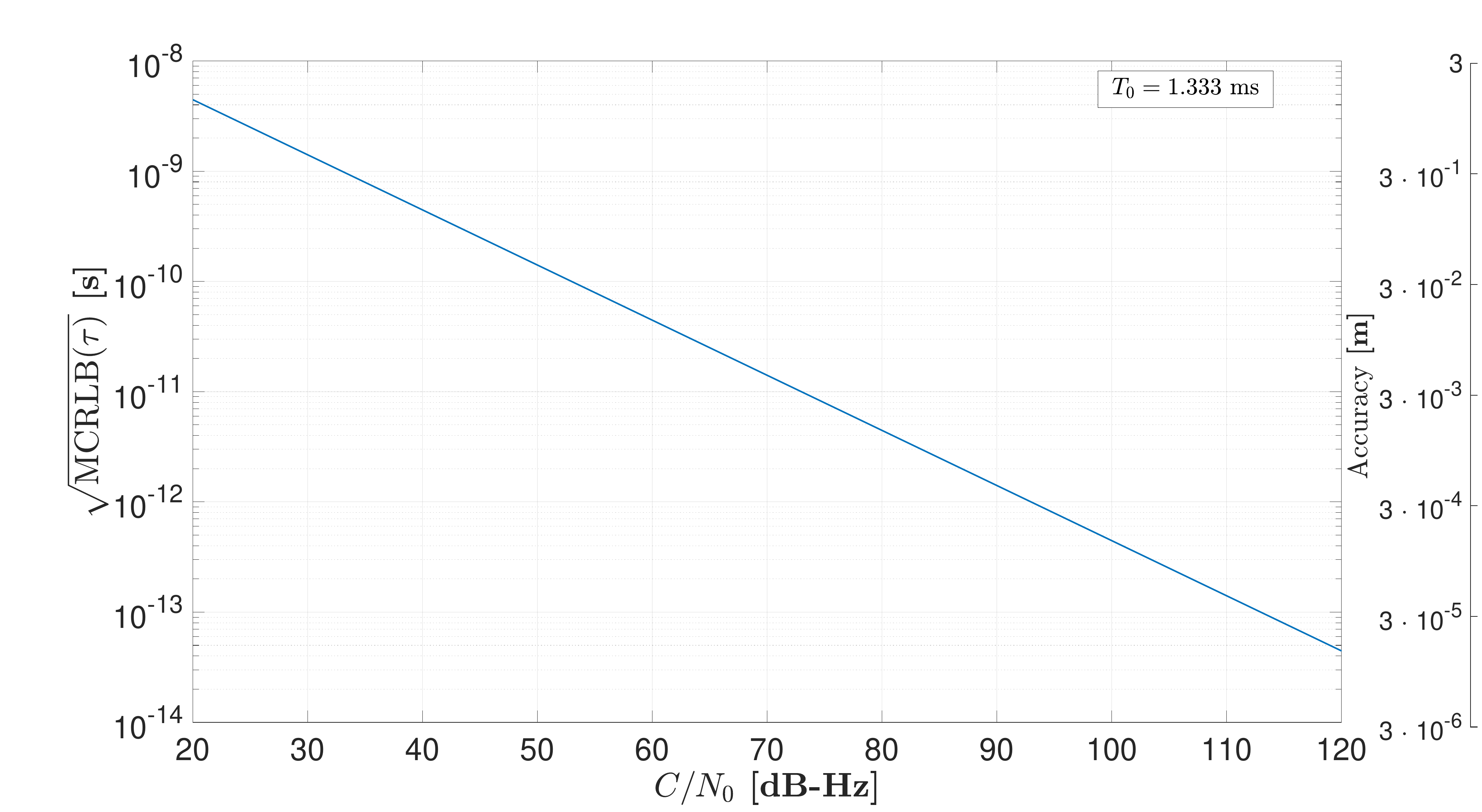}}
\subfloat[Frequency Shift, $\upsilon$. \label{fig:mcrb_v}]{\includegraphics[width=0.5\columnwidth]{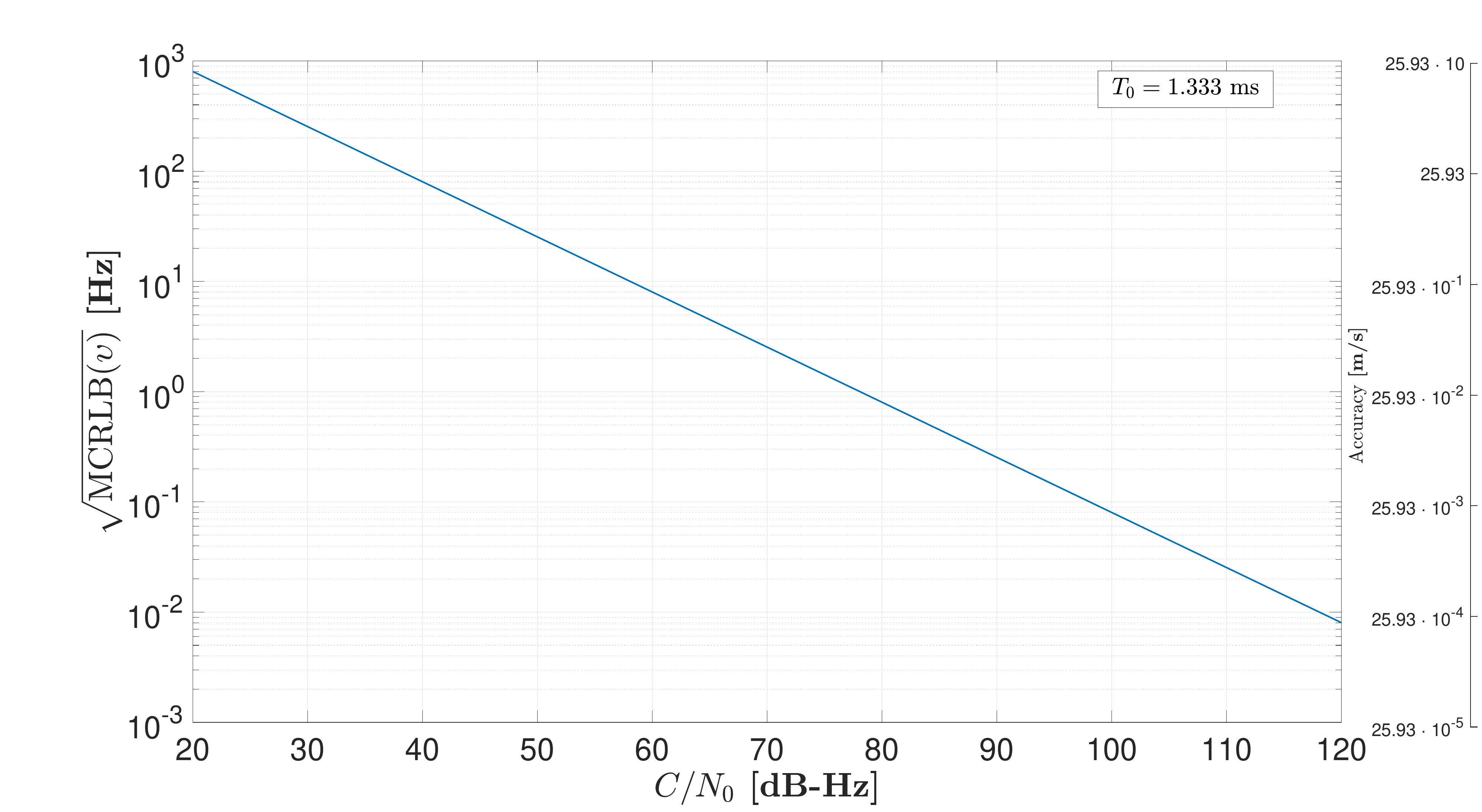}}\\
\subfloat[Phase Delay, $\theta$. \label{fig:mcrb_theta}]{\includegraphics[width=0.5\columnwidth]{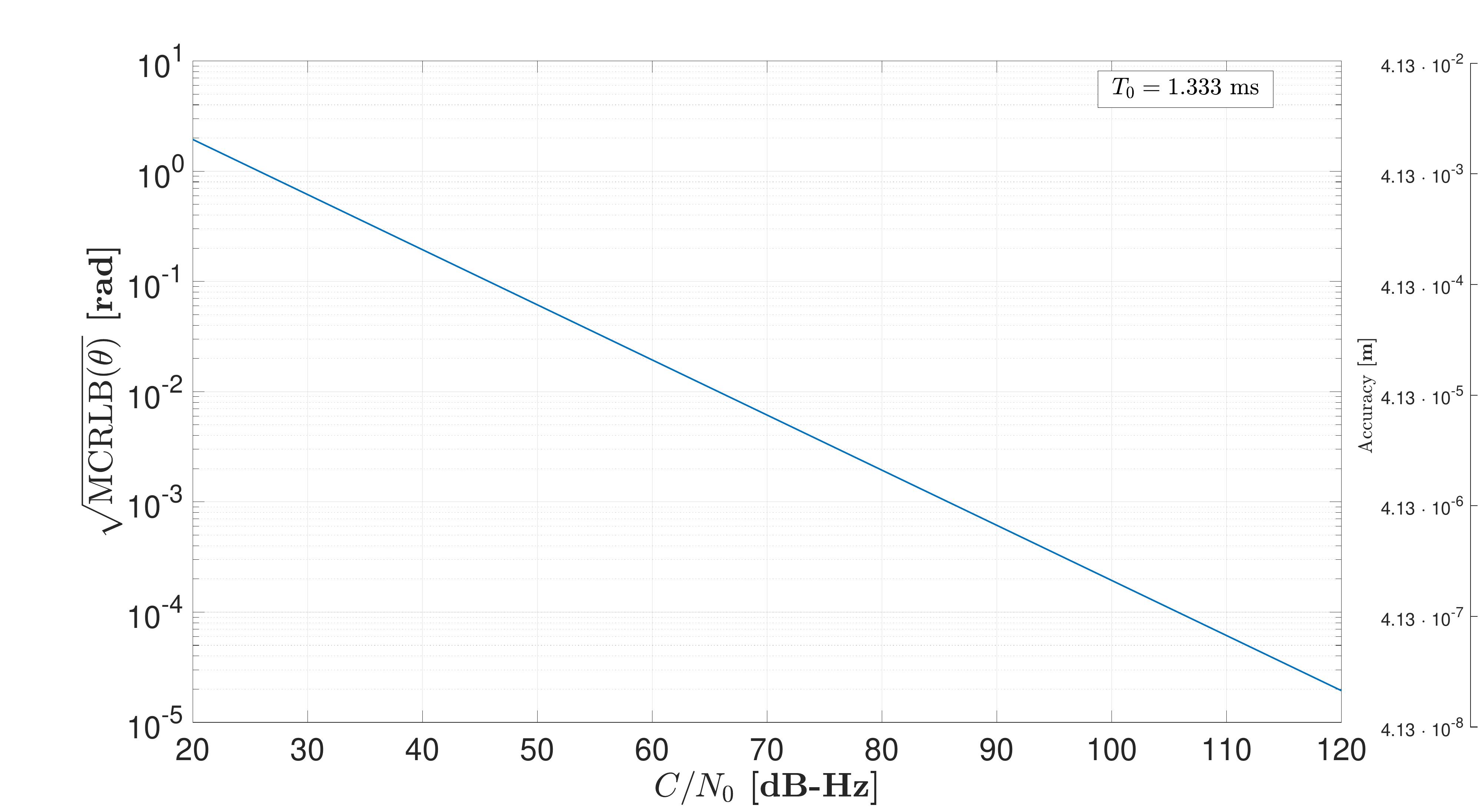}}
\subfloat[Angle of Arrival $\beta$.\label{fig:mcrb_aoa}]{\includegraphics[width=0.5\columnwidth]{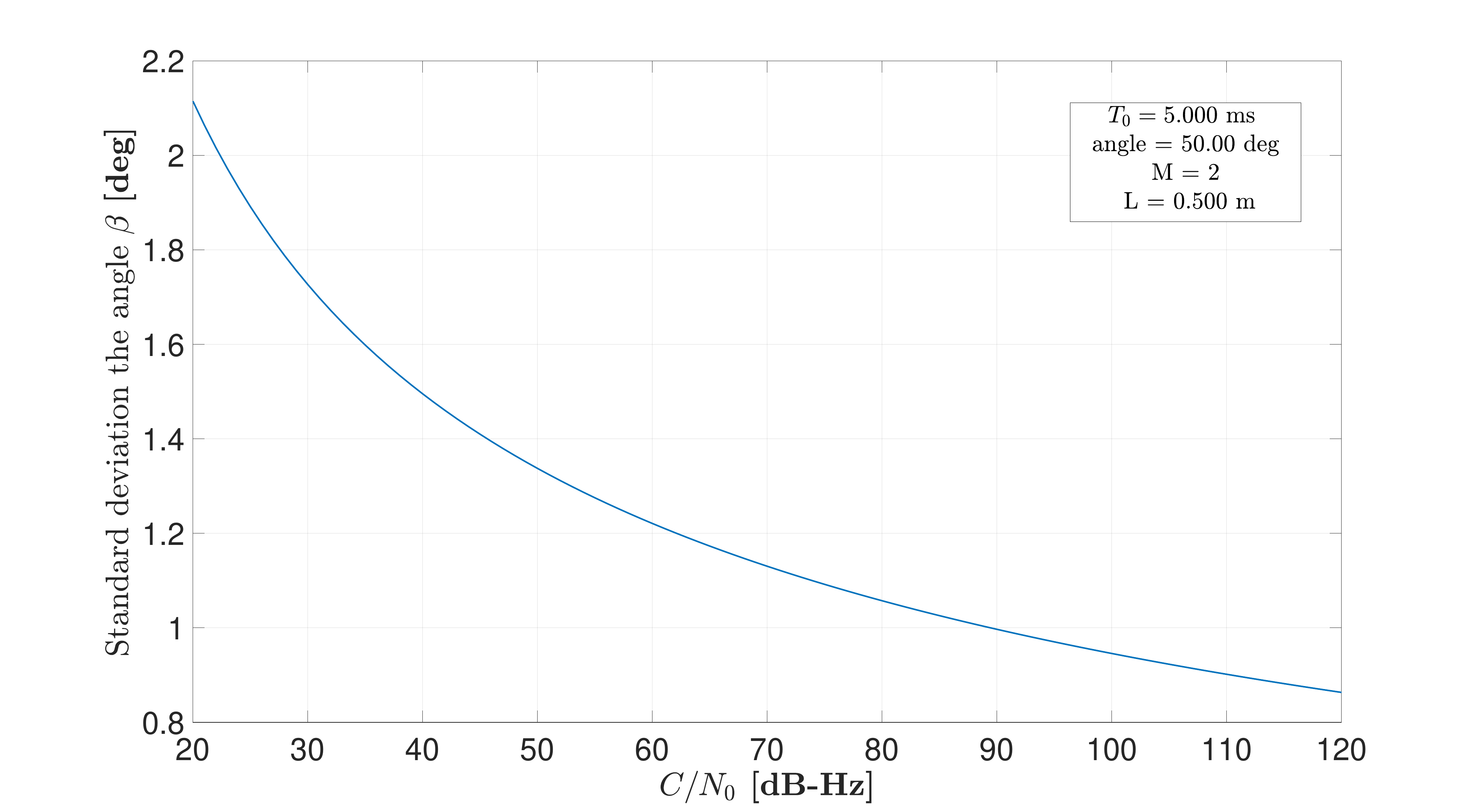}}
\caption{\ac{mcrlb} vs \ac{cn0} for the measurement under consideration considering Starlink as \ac{soop}.}
\label{fig:mcrb_cn0}
\end{figure}

In Fig. \ref{fig:mcrb_cn0} are represented the \ac{mcrlb} of the propagation time, frequency offset, phase shift, and angle of arrival for Starlink signal, as computed in \eqref{mcrlb_tau_cndr}, \eqref{mcrlb_theta_cndr}, \eqref{mcrlb_upsilon_cndr} and \eqref{eq:mcrlb_aoa}, respectively.
In Figs. \ref{fig:mcrb_tau} - \ref{fig:mcrb_theta}, the observation time is equal to the duration of a frame, while in Fig. \ref{fig:mcrb_aoa} it is fixed to \SI{5}{\milli\second}. In particular, for the latter, we consider an angle of \SI{50}{\degree}, $M=2$ antennas, and array length of $L=\SI{0.5}{\meter}$. The conversions on the right axes were made using $c$ as propagation speed and $f_{\mathrm{c}} = 11.57$ GHz as carrier frequency.

\begin{figure}[ht]
\centering
\subfloat[\label{fig:mcrb_tau_T0}]{\includegraphics[width=0.5\columnwidth]{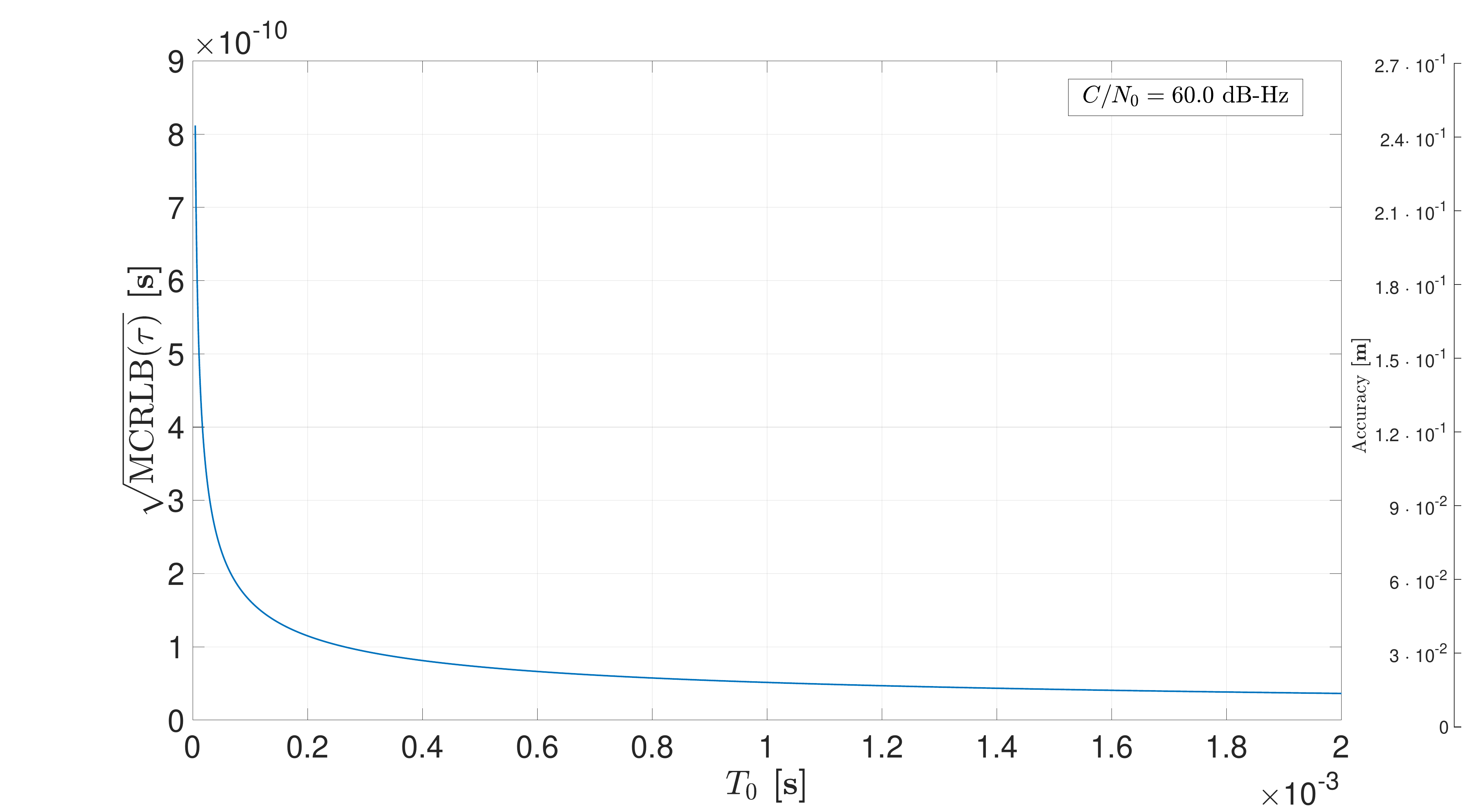}}
\subfloat[\label{fig:mcrb_v_T0}]{\includegraphics[width=0.5\columnwidth]{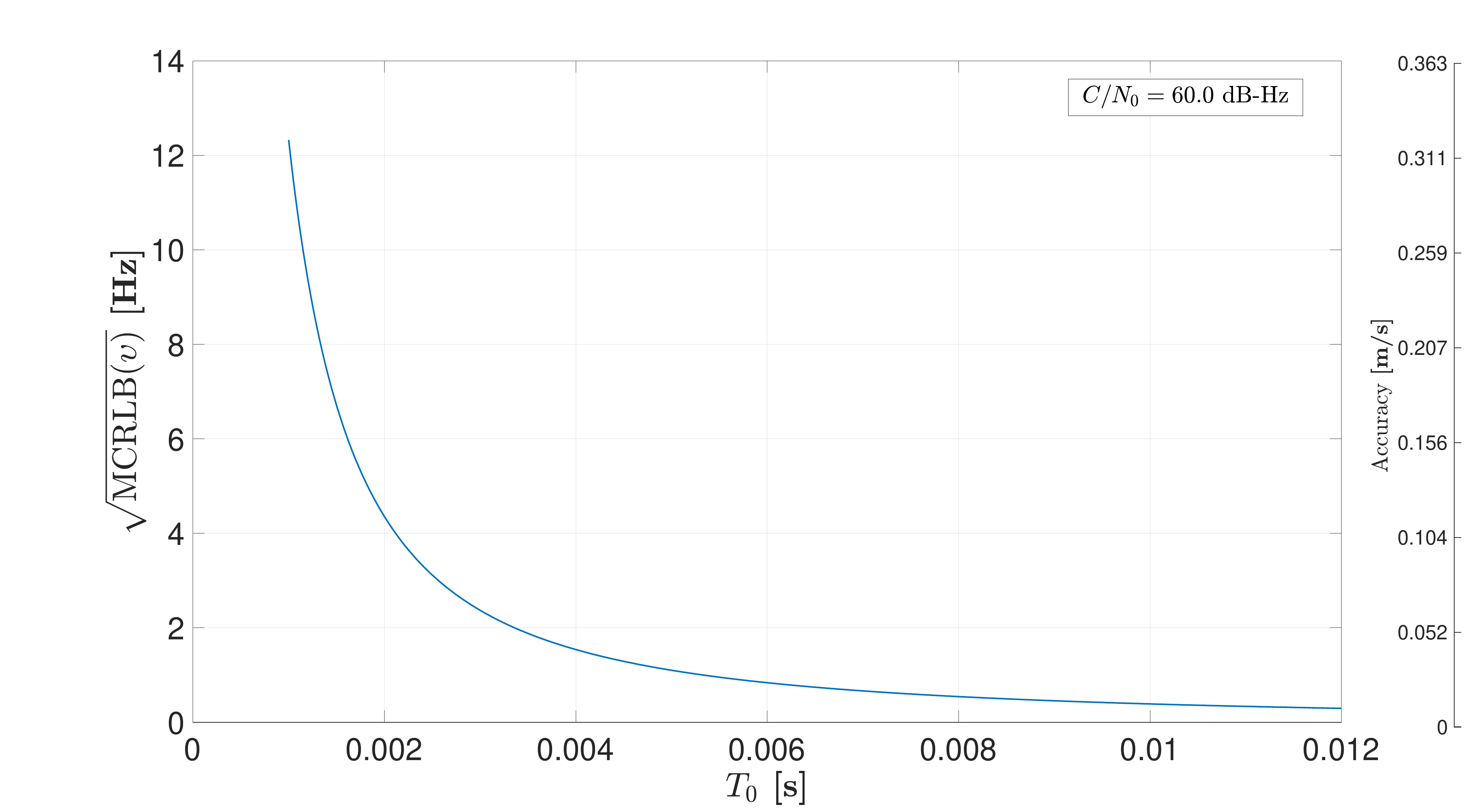}}\\
\subfloat[\label{fig:mcrb_theta_T0}]{\includegraphics[width=0.5\columnwidth]{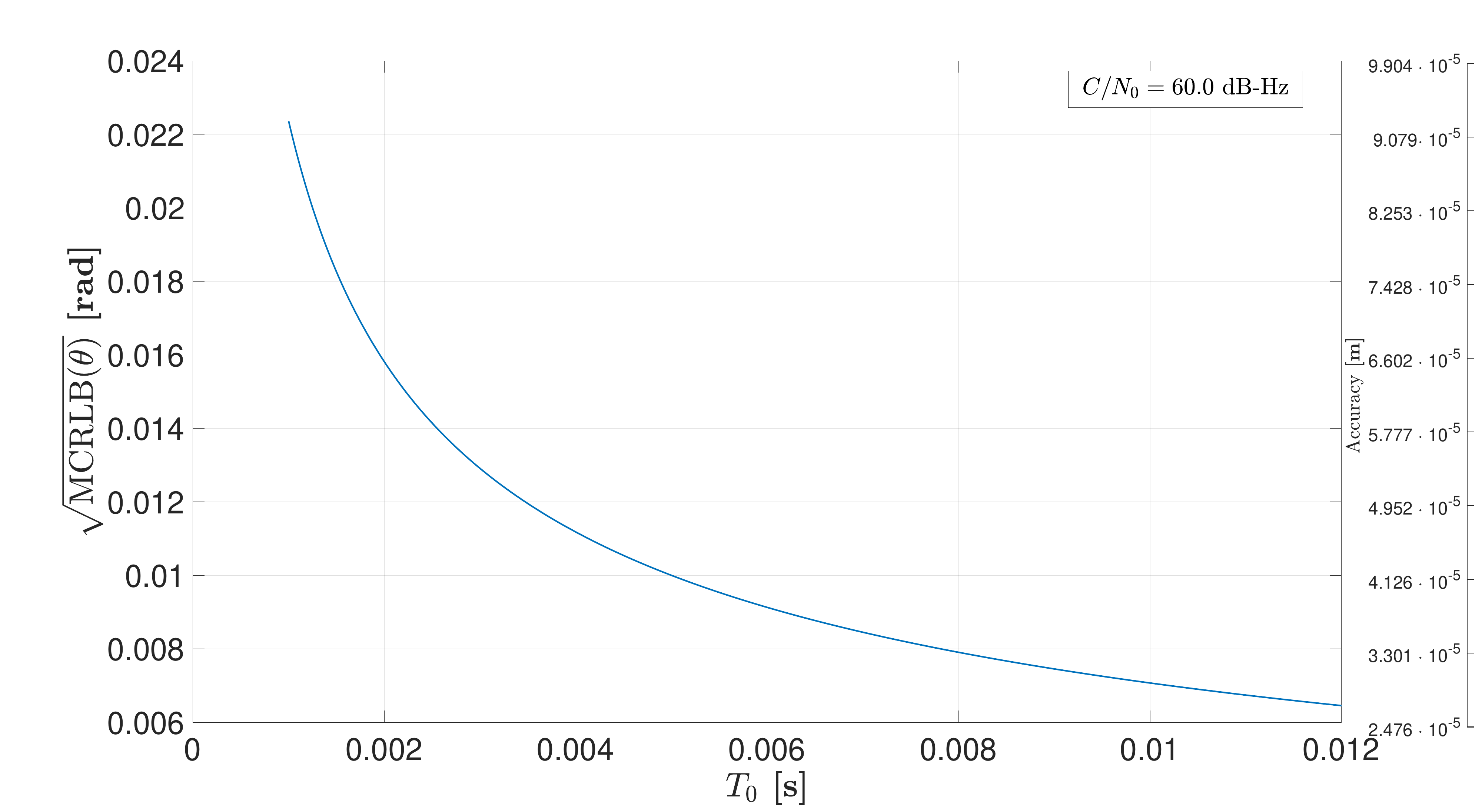}}
\caption{\ac{mcrlb} vs the observation time $T_0$ for (a) $\tau$, (b) $\upsilon$, and (c) $\theta$.}
\label{fig:mcrb_cn0_T0}
\end{figure}

Fig. \ref{fig:mcrb_cn0_T0} reports the \acp{mcrlb} for the same parameter keeping $C/N_0$ fixed to 60 dB-Hz and varying the observation time $T_0$. As in the first batch, we see that the \ac{mcrlb} decreases inversely with the parameters under observation. 

\begin{figure}[ht]
\centering
\subfloat[Propagation Delay, $\tau$]{\includegraphics[width=0.5\columnwidth]{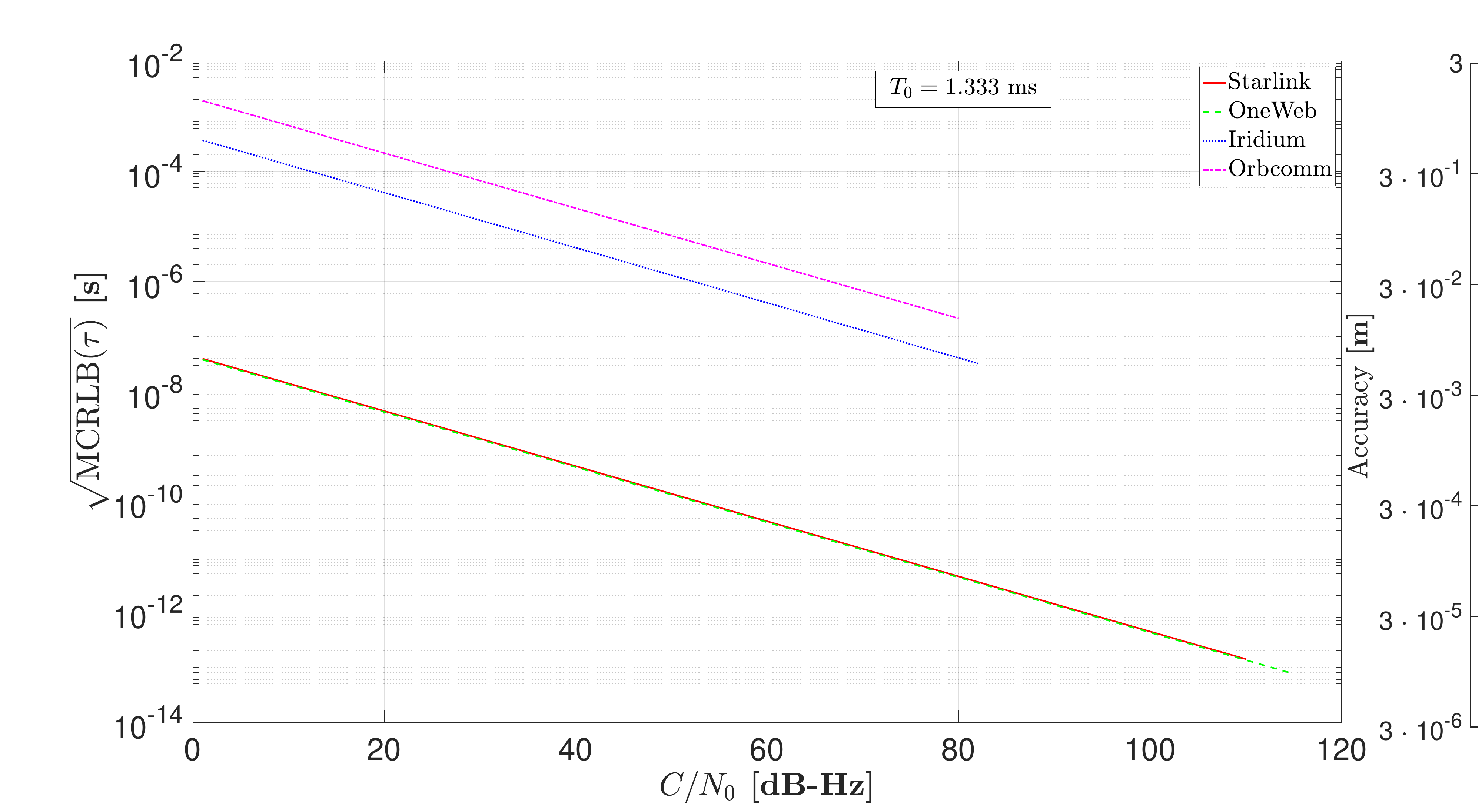}}
\subfloat[Frequency Shift, $\upsilon$]{\includegraphics[width=0.5\columnwidth]{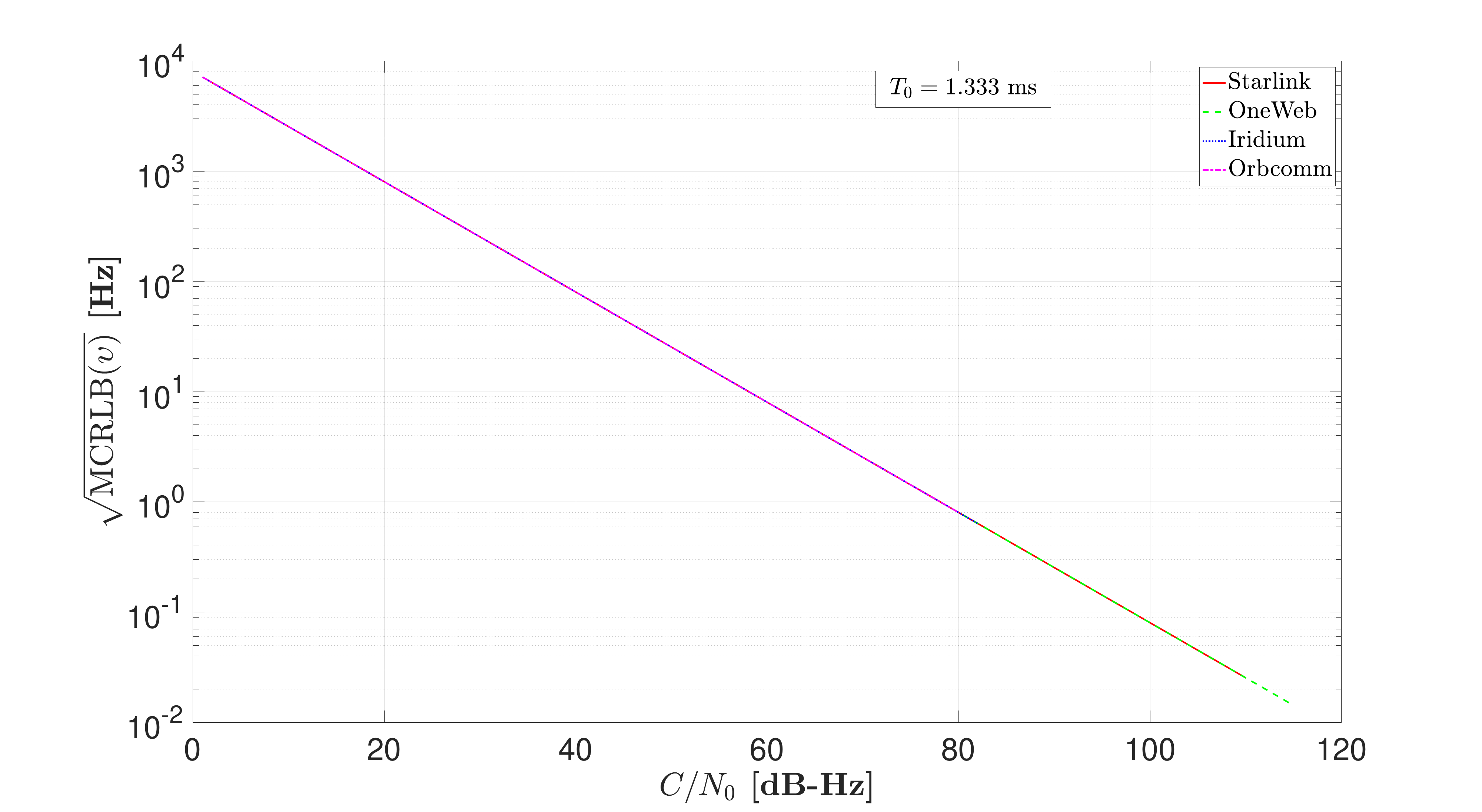}}\\
\subfloat[Phase Delay, $\theta$]{\includegraphics[width=0.5\columnwidth]{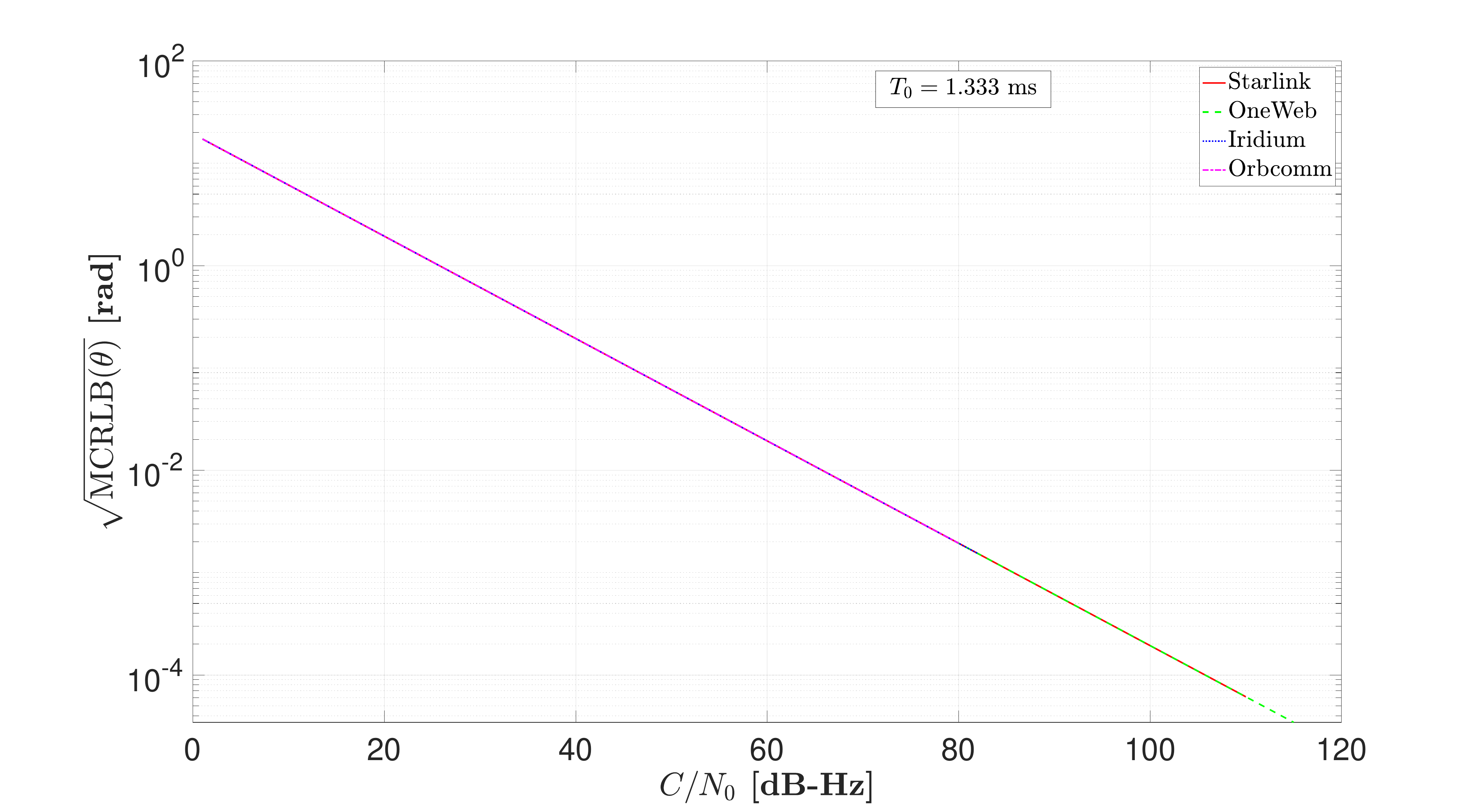}}
\caption{\ac{mcrlb} vs the observation time \ac{cn0} for (a) $\tau$, (b) $\upsilon$, and (c) $\theta$ considering OrbComm, Iridium, Starlink and OneWeb, with $T_0=\SI{1.33}{\milli\second}$.}
\label{fig:mcrlb_confronti}
\end{figure}

Finally Fig. \ref{fig:mcrlb_confronti} reports the \acp{mcrlb} as a function of the \ac{cn0} for a fixed $T_0 = \SI{1.33}{\milli\second}$, corresponding to the frame of a Starlink frame. In particular, we consider a single antenna receiver, thus the \ac{aoa} is not a viable measurement.
While the results are comparable for phase and frequency shift, the best performance for the propagation delay is achieved by Starlink and OneWeb. Indeed, the propagation delay is a valid candidate, due to the easier ambiguity resolution. 

\subsection{MCRLB validation under realistic conditions}
To task the validity of the proposed model we run a simulation using the OFDM Starlink signal simulator published in \cite{komodromos2023IONsimulator}. In particular, we compare the accuracy that characterizes the estimated propagation delay $\tau$ and its \ac{mcrlb} as a function of $C/N_0$. For the acquisition process, we used the whole channel bandwidth of \SI{240}{\mega \hertz}. The procedure is similar to the one used for \ac{gnss}. However, since no \ac{prn} sequence is available, cross-correlation was performed making use of the two known synchronization sequences, the \ac{pss} and \ac{sss}, that are transmitted by Starlink satellites at the beginning of each frame. The total duration of these two \ac{ofdm} symbols is \SI{8.8}{\micro\second}.

\begin{figure}[ht]
\centering
\includegraphics[width=0.5\columnwidth]{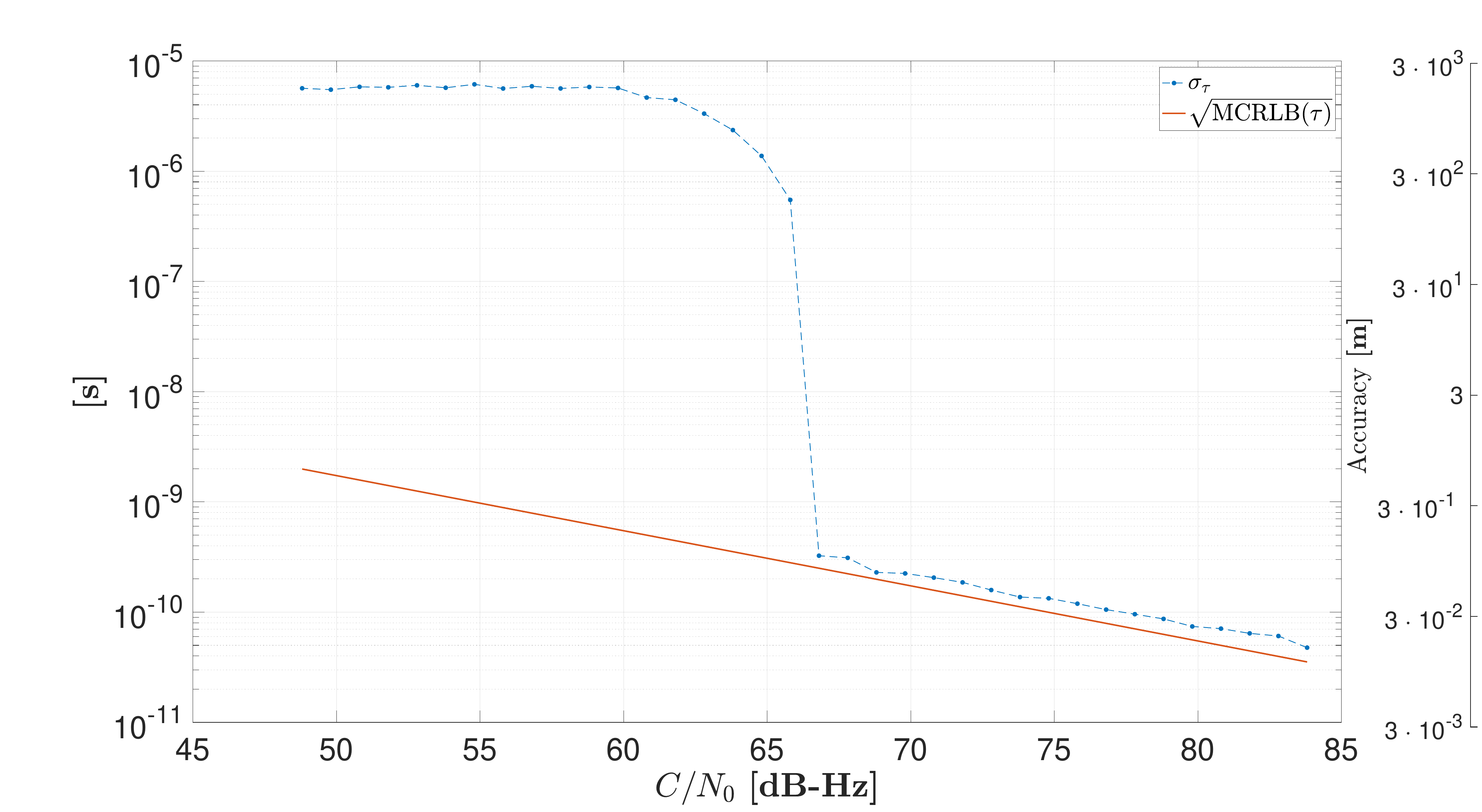}
\caption{Acquisition results of Starlink OFDM and Modified Cramér-Rao Lower Bound as function of $C/N_0$}
\label{fig:std_tau}
\end{figure}

We performed 300 simulations for each $C/N_0$ value, with a time interval of \SI{20.83}{\micro\second}. The plot in Fig. \ref{fig:std_tau} shows that for values of $C/N_0$ above $66$ dB-Hz, sub-meter accuracy can be achieved while below this threshold the results deviate from the \ac{mcrlb} and tend to have a uniform distribution within the search interval. 

\section{Geometry and Availability Considerations}
In this Section, we investigate the performance limits of the \ac{soop}-aided navigation due to availability and geometry, i.e., the placement of the satellites with respect to the receiver position.

To do so we consider two metrics. First, we consider the \ac{ccdf} of the number of satellites in view: indeed, while we are interested in having at least 4 satellites in view for most of the time, having a high number of satellites typically leads to a more accurate \ac{pvt} solution. 
Next, we consider the \ac{cdf} of the \ac{gdop}, as the \ac{gdop} also reports information about how spread out the satellites are with respect to the receiver, with sparser constellation having a lower \ac{gdop}. Then, the derivation of the \ac{gdop} can be used to derive the overall positioning accuracy by using \eqref{eq:accuracy_GDOPUERE}.

We consider two scenarios, which we call Scenario 1 and 2, and run a Montecarlo simulation for each, where the satellites' positions are obtained via an orbital propagator with used real-world broadcast ephemeris.
In both scenarios, we considered as representative locations Padova, the Svalbard islands, La Reunion island, and ESTEC.
In Scenario 1 we neglected the visibility effect induced by the beamwidth and masking angle, thus considering $\varphi = \SI{90}{\degree}$ and $\theta = \SI{10}{\degree}$, as typically chosen for \ac{meo}-GNSS navigation. On the other hand, in Scenario 2, we consider a $\varphi = [\SI{30}{\degree}, \ldots, \SI{80.92}{\degree}]$ and $\theta = \SI{40}{\degree}$.

\paragraph{Analysis for Scenario 1}

The parameters of Scenario 1 are reported in Tab. \ref{tab:simParam1}. For this first set of simulations, we neglected the visibility effect induced by the beamwidth by considering $\varphi = \SI{90}{\degree}$. The impact of beamwidth will be analyzed instead in Scenario 2. 

\begin{table}[ht]
    \renewcommand{\arraystretch}{1.2}
	\centering
 	\caption{List of simulation parameters for Scenario 1.}\label{tab:simParam1}
	\begin{tabular}{|l|c|}
 \hline
		\textbf{Constellations} & Orbcomm, Iridium, Starlink, OneWeb  \\ \hline
            \multirow{4}{*}{\textbf{\textbf{Receiver Locations}}}  & Padova, Italy [\SI{45.409}{\degree} N, \SI{11.894}{\degree} E]\\
            & Longyearbyen, Svalbard [\SI{78.224}{\degree} N, \SI{15.637}{\degree} E]\\
            & ESTEC, the Netherlands  [\SI{52.219}{\degree} N, \SI{4.419}{\degree} E ]\\
            & Saint Denis, La Reunion  [\SI{20.883}{\degree} S, \SI{55.450}{\degree} E]\\ \hline
            \textbf{Time} & April $19$, $2024$, $00:00$ - $23:59$	\\ \hline
		\textbf{Beamwidth}, $\varphi$ & \SI{90}{\degree} \\ \hline
		\textbf{Masking Angle}, $\theta$ & \SI{10}{\degree}\\ \hline
	\end{tabular}
\end{table}

\begin{figure}[ht]
    \centering
    \subfloat[Iridium.\label{fig:iridiumNext_CCDF}]{\includegraphics[width=0.4\columnwidth]{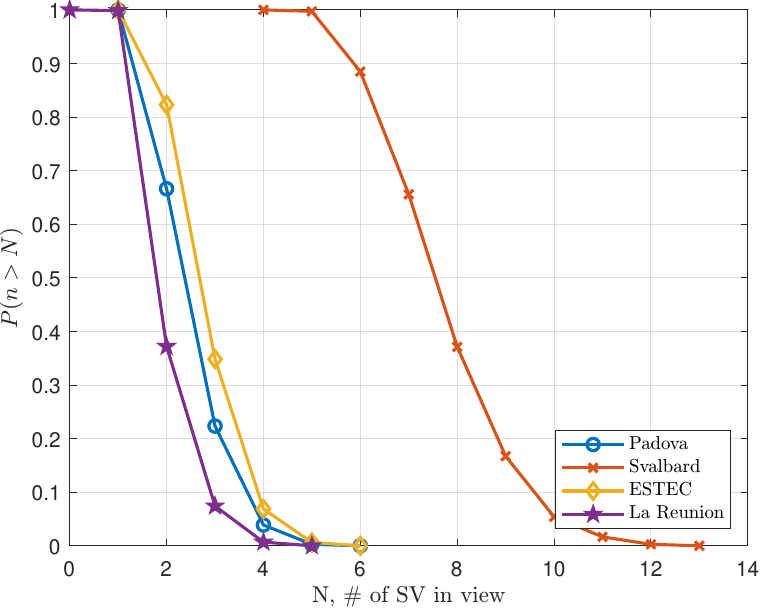}}
    \subfloat[ORBCOMM.\label{fig:orbcomm_CCDF}]{\includegraphics[width=0.4\columnwidth]{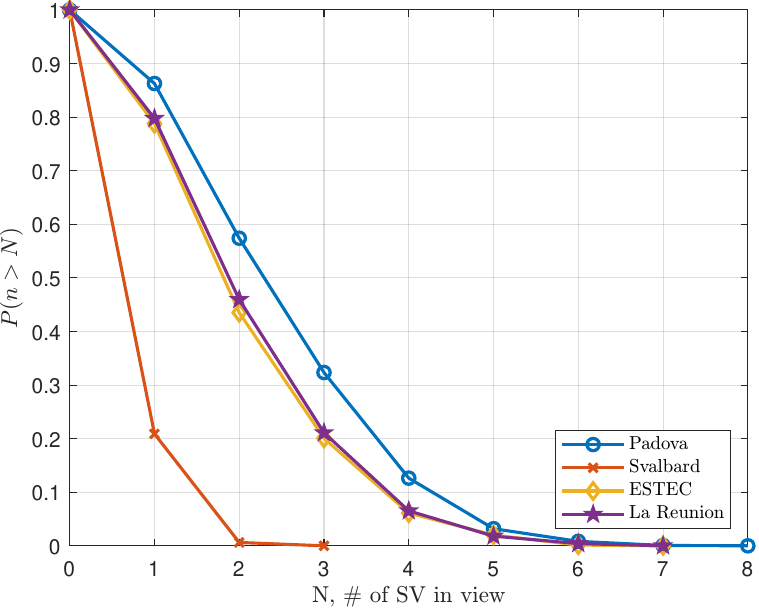}}\\
    \subfloat[OneWeb.\label{fig:oneweb_CCDF}]{\includegraphics[width=0.4\columnwidth]{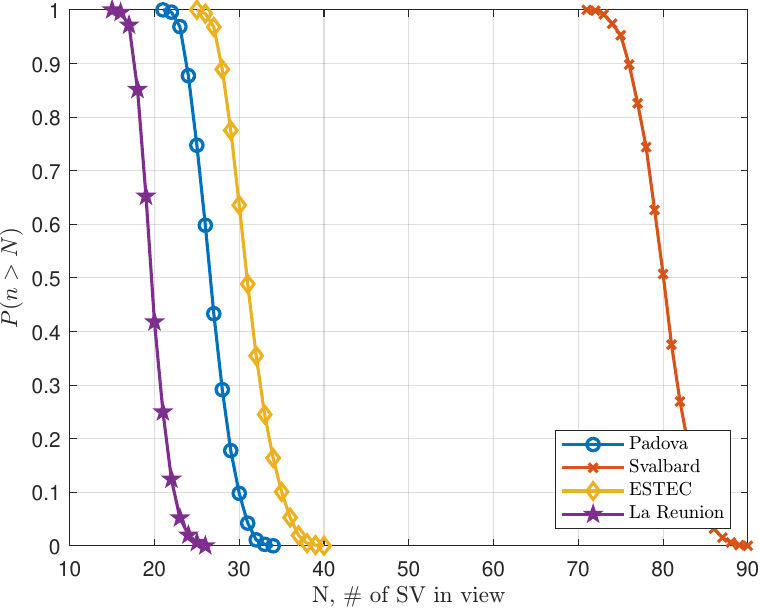}}
    \subfloat[Starlink.\label{fig:starlink_CCDF}]{\includegraphics[width=0.4\columnwidth]{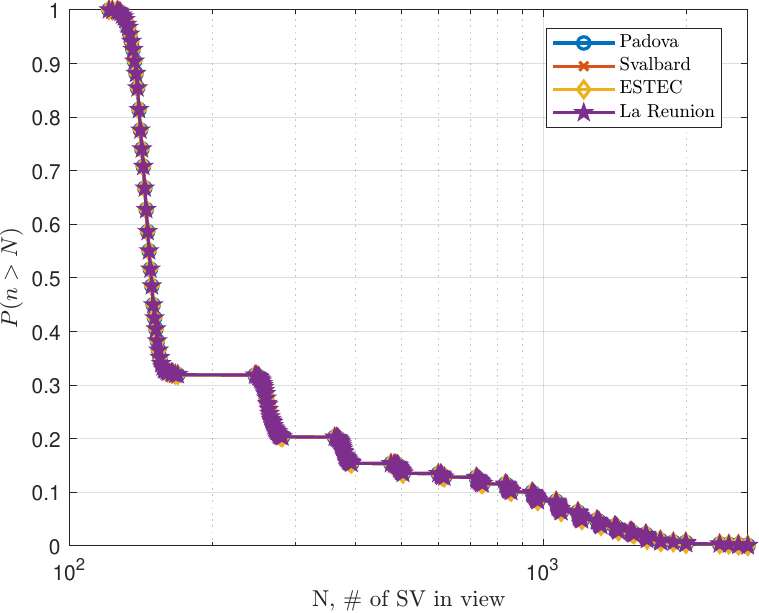}}
    \caption{\Acp{ccdf} of the number of satellites in view, considering the scenario described in Tab. \ref{tab:simParam1}.}\label{fig:ccdfScenario1}
\end{figure}

Fig. \ref{fig:ccdfScenario1} reports the \acp{ccdf} of the number of satellites in view during the observation period computed for Scenario 1. As expected, due to the larger size of the constellation the best performance is achieved by Starlink and OneWeb which both always guarantee at least $20$ satellites in view from all the considered receiver locations. Interestingly, we note that, due to higher orbit inclination differently from Starlink, OneWeb achieves better performance at higher-latitude positions.

\begin{figure}
    \centering
    \subfloat[Iridium-NEXT.\label{fig:iridium-NEXT_GDOP_CDF}]{\includegraphics[width=0.4\columnwidth]{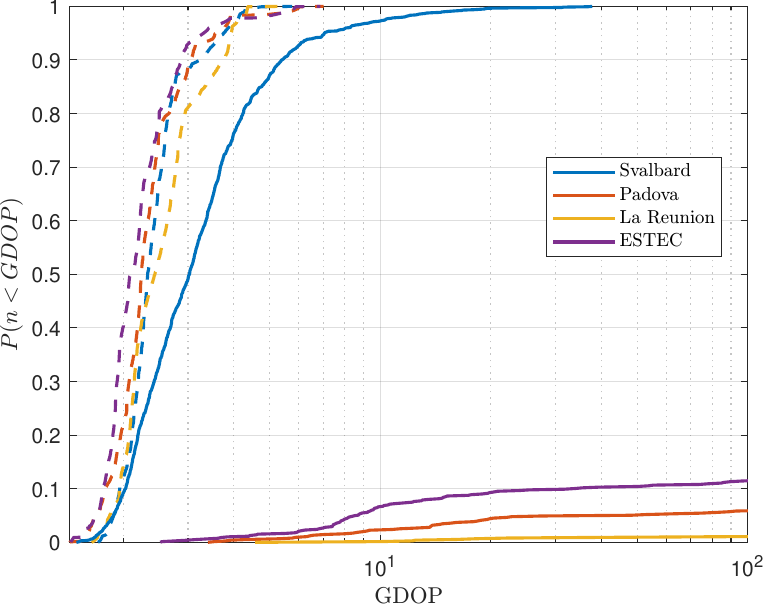}}
    \subfloat[ORBCOMM.\label{fig:orbcomm_GDOP_CDF}]{\includegraphics[width=0.4\columnwidth]{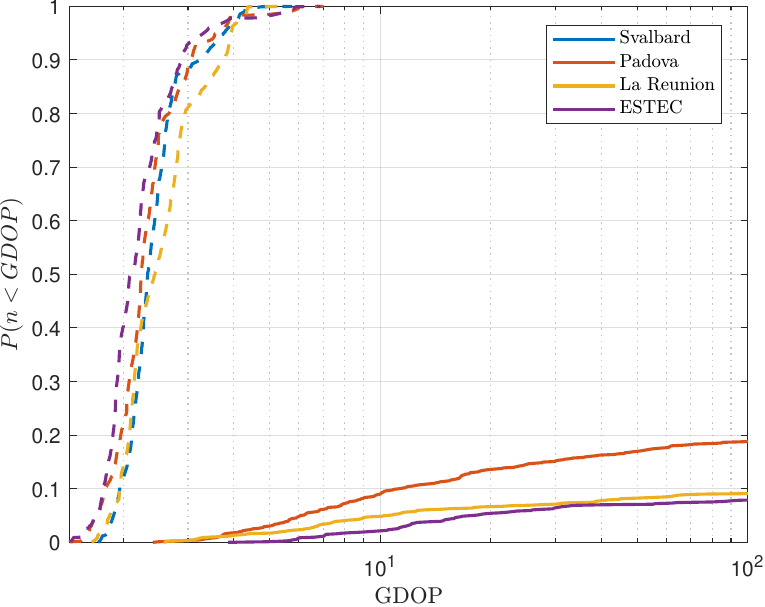}}\\
    \subfloat[OneWeb.\label{fig:oneweb_GDOP_CDF}]{\includegraphics[width=0.4\columnwidth]{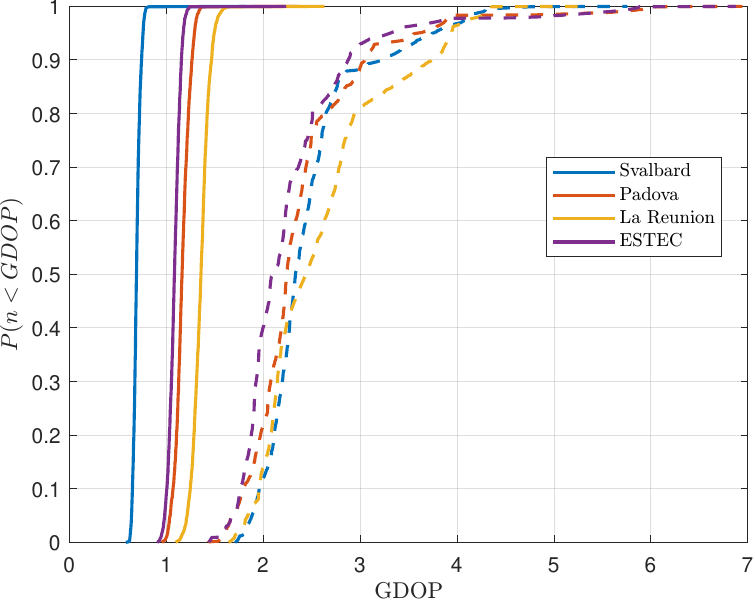}}
    \subfloat[Starlink.\label{fig:starlink_GDOP_CDF}]{\includegraphics[width=0.4\columnwidth]{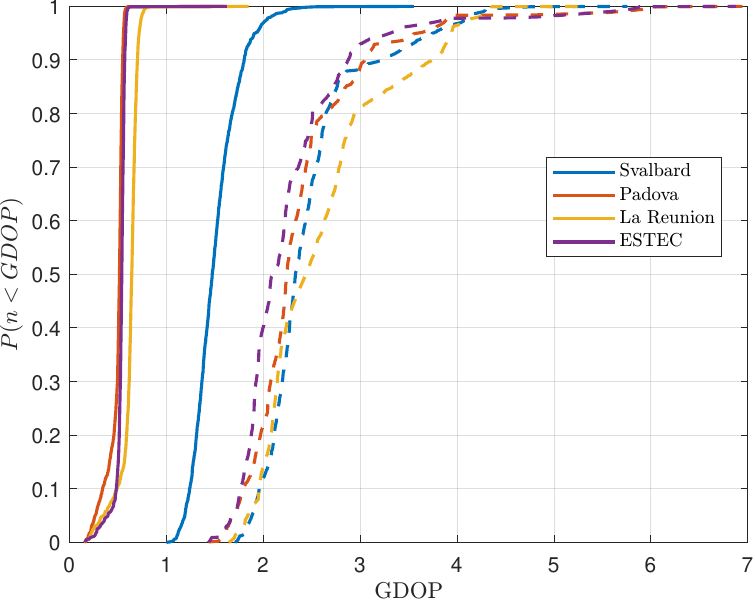}}
    \caption{\Acp{gdop} of the number of satellites in view, considering the scenario described in Tab. \ref{tab:simParam1}, and comparison with Galileo dashed).}\label{fig:GDOP_scenario1}
\end{figure}

Fig. \ref{fig:GDOP_scenario1} reports the \ac{cdf} of the \acp{gdop} computed for Scenario 1 compared with the one achieved instead by using Galileo. The results mirror the one achieved in Fig. \ref{fig:ccdfScenario1}, with Starlink and OneWeb achieving the best performance due to their larger constellation size and even outperforming Galileo. Additionally, we notice again that OneWeb achieves better performance than Starlink from high-latitude locations.

The results concerning the \ac{gdop} are summarized in Tab. \ref{tab:GDOPScenario1} where we reported the mean and \ac{std} of the measured \ac{gdop}.

\begin{table}
    \caption{Average and \ac{std} of the \ac{gdop} for the different constellations, computed over the parameters reported in Tab. \ref{tab:simParam1}. In red the systems which outperform Galileo. With '/' we indicated that no \ac{pvt} could be computed and thus no \ac{gdop} can be reported.}\label{tab:GDOPScenario1}
    \centering
    \renewcommand*{\arraystretch}{1.1}
    \begin{tabular}{|c|cccc|}
			\hline
                & \multicolumn{4}{c|}{\textbf{Receiver Locations}} \\
			\textbf{Constellation} &   \textbf{Svalbard}           &  \textbf{Padova}           &  \textbf{La Reunion}            &    \textbf{ESTEC}         \\ \hline
			Iridium NEXT & $3.608 \pm 2.557$ & / & /  & / \\		
			OrbComm            & /   & /   & /   & /   \\
			\color{UniPDred}OneWeb             & $\color{UniPDred}0.692 \pm 0.035$ & $\color{UniPDred}1.162 \pm 0.074$ & $\color{UniPDred}1.353 \pm 0.085$ & $\color{UniPDred}1.079 \pm 0.058$ \\
			\color{UniPDred}Starlink           & $\color{UniPDred}1.500 \pm 0.227$ & $\color{UniPDred}0.482 \pm 0.086$ & $\color{UniPDred}0.614  \pm 0.108$ & $\color{UniPDred}0.523 \pm 0.064$ \\
			\hline
			\textbf{{Galileo}}            & $\bm{2.465 \pm 0.523}$ & $\bm{2.380 \pm 0.634}$ & $\bm{2.589 \pm 0.640}$ & $\bm{2.661 \pm 0.635}$ \\
			\hline			
	\end{tabular}
    
\end{table}

\paragraph{Analysis for Scenario 2}
In order to compare the performance of OneWeb and Starlink, we now consider the stricter Scenario 2, whose parameters are reported in \ref{tab:simParam2}, where the parameters for beamwidth $\varphi$  the masking angle $\theta$ follow the descriptions reported in the proposals \cite{starlinkTISS,oneWebTISS}.
In particular, a more narrow masking angle $\theta = \SI{40}{\degree}$ and beamwidth $\varphi = \SI{30}{\degree}, \SI{40}{\degree}, \ldots, \SI{80.92}{\degree}$ are considered, where $\varphi_\mathrm{max} =  \SI{80.92}{\degree}$ is the maximum beamwidth used by Starlink \cite{starlinkTISS}.
Indeed, since the \ac{gdop} increases with the number of satellites in view, for the sake of conciseness, we will only report the \ac{ccdf} of the number of satellites in view. 

We remark that OneWeb \acp{sv} are designed to have a dynamic beam able to limit overlap between spots of different \acp{sv} \cite{oneWebTISS}, therefore the analysis can considered to be a best case for OneWeb.

\begin{table}
	\caption{List of simulation parameters for Scenario 2.}\label{tab:simParam2}
	\centering
	\begin{tabular}{|l|c|}
		\hline
		\textbf{Constellations} & Starlink, OneWeb\\
		\textbf{Time} & April 19, 2024, 00:00 - 23:59	\\
            \multirow{4}{*}{\textbf{\textbf{Receiver Locations}}}  & Padova, Italy [\SI{45.409}{\degree} N, \SI{11.894}{\degree} E]\\
            & Longyearbyen, Svalbard [\SI{78.224}{\degree} N, \SI{15.637}{\degree} E]\\
            & ESTEC, The Netherlands  [\SI{52.219}{\degree} N, \SI{4.419}{\degree} E ]\\
            & Saint Denis, La Reunion  [\SI{20.883}{\degree} S, \SI{55.450}{\degree} E]\\ \hline
            \textbf{Beamwidth}, $\varphi$ &  \SI{30}{\degree},  \SI{40}{\degree},  \SI{50}{\degree},  \SI{60}{\degree},  \SI{70}{\degree}, \SI{80.92}{\degree} \\
		\textbf{Masking Angle}, $\phi$ & \SI{40}{\degree}\\ \hline
	\end{tabular}
\end{table}

\begin{figure}
	\setlength{\figurewidth}{0.4\columnwidth}
	\setlength{\figureheight}{0.8\figurewidth}
	\centering

\definecolor{mycolor1}{RGB}{215,48,39}%
\definecolor{mycolor2}{RGB}{252,141,89}%
\definecolor{mycolor3}{RGB}{254,224,139}%
\definecolor{mycolor4}{RGB}{217,239,139}%
\definecolor{mycolor5}{RGB}{145,207,96}%
\definecolor{mycolor6}{RGB}{26,152,80}%

\begin{tikzpicture}

\begin{axis}[%
width=0.951\figurewidth,
height=\figureheight,
scale only axis,
xmin=0,
xmax=30,
xlabel style={font=\color{white!15!black}},
xlabel={N, \# of SV in view},
ymin=0,
ymax=1,
ylabel style={font=\color{white!15!black}},
ylabel={$P(n>N)$},
ytick distance=0.1,
axis background/.style={fill=white},
xmajorgrids,
ymajorgrids,
enlargelimits=false,xlabel style={font=\scriptsize},ylabel style={font=\scriptsize},ticklabel style={font=\scriptsize},
legend style={at={(0.97,0.97)}, anchor=north east, font=\scriptsize, legend cell align=center, align=left, draw=white!15!black}
]
\addplot [color=mycolor1, line width=1.5pt, mark=o, mark options={solid, mycolor1}]
  table[row sep=crcr]{%
0	1\\
1	0.582234559333796\\
2	0.179736294240112\\
3	0.0284524635669676\\
4	0.00346981263011781\\
5	0.000693962526023384\\
6	0\\
};
\addlegendentry{$\varphi = \SI{30}{\degree}$}
\addplot [color=mycolor2, line width=1.5pt, mark=o, mark options={solid, mycolor2}]
  table[row sep=crcr]{%
0	1\\
1	0.848716169326856\\
2	0.472588480222068\\
3	0.145732130464955\\
4	0.0346981263011799\\
5	0.00693962526023562\\
6	0.00277585010409442\\
7	0\\
};
\addlegendentry{$\varphi = \SI{40}{\degree}$}
\addplot [color=mycolor3, line width=1.5pt, mark=o, mark options={solid, mycolor3}]
  table[row sep=crcr]{%
0	1\\
1	0.984038861901457\\
2	0.834836918806385\\
3	0.498959056210964\\
4	0.20263705759889\\
5	0.0610687022900755\\
6	0.0145732130464964\\
7	0.00416377515614208\\
8	0\\
};
\addlegendentry{$\varphi = \SI{50}{\degree}$}
\addplot [color=mycolor4, line width=1.5pt, mark=o, mark options={solid, mycolor4}]
  table[row sep=crcr]{%
1	1\\
2	0.995836224843858\\
3	0.945176960444137\\
4	0.770992366412214\\
5	0.524635669673838\\
6	0.271339347675225\\
7	0.126995142262318\\
8	0.0499653018736996\\
9	0.0201249132546835\\
10	0.00277585010409531\\
11	0\\
};
\addlegendentry{$\varphi = \SI{60}{\degree}$}
\addplot [color=mycolor5, line width=1.5pt, mark=o, mark options={solid, mycolor5}]
  table[row sep=crcr]{%
3	1\\
4	0.993060374739763\\
5	0.957668285912561\\
6	0.861901457321306\\
7	0.710617626648162\\
8	0.489243580846633\\
9	0.298403886190147\\
10	0.155447605829284\\
11	0.0721721027064532\\
12	0.0346981263011799\\
13	0.0104094378903525\\
14	0.00277585010409354\\
15	0.00069396252602516\\
16	0\\
};
\addlegendentry{$\varphi = \SI{70}{\degree}$}
\addplot [color=mycolor6, line width=1.5pt, mark=o, mark options={solid, mycolor6}]
  table[row sep=crcr]{%
6	1\\
7	0.998612074947953\\
8	0.989590562109647\\
9	0.958362248438583\\
10	0.870922970159612\\
11	0.74739764052741\\
12	0.562109646079112\\
13	0.390700902151284\\
14	0.249826509368493\\
15	0.147120055517004\\
16	0.0770298403886187\\
17	0.0360860513532266\\
18	0.0208188757807086\\
19	0.00485773768216546\\
20	0.00138792505204677\\
21	0.00069396252602516\\
22	0\\
};
\addlegendentry{$\varphi = \SI{80.92}{\degree}$}
\end{axis}
\end{tikzpicture}%
	\caption{\ac{ccdf} of the number of Starlink \acp{sv} in view, for different beam width angles $\varphi$, with scenario described in Tab. \ref{tab:simParam2} (Padova only).}
	\label{fig:starlinkBeam}	
\end{figure}
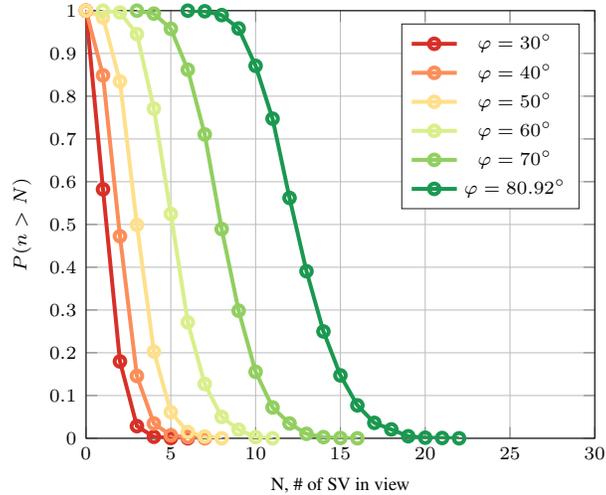

The results for the Montecarlo simulations for OneWeb and Starlink concerning the availability are reported in Figs. \ref{fig:oneWEbBeam} and \ref{fig:starlinkBeam}, respectively. 
For Starlink, only the results from Padova are reported since, as shown from the previous batch of results, no relevant difference is observed from one location to the other. 
Concerning OneWeb, following the trend shown in the previous results, due to the high orbit inclination, the 
best results are measured from the Svalbard islands, from which it is also possible to outperform the performance obtained instead by Starlink. 
On the other hand, when considering lower latitude locations, Starlink achieves much better performance than OneWeb, for all the considered beamwidths. 
More in detail, numerical results show that except from the Svalbard, even considering $\theta_\mathrm{max}$ a \ac{pvt} can be computed using only OneWeb satellites at most the $76\%$ of the time (Fig. \ref{fig:oneWebBeamPadova}). On the other hand, when using Starlink, the same percentage is achieved using a much narrower beamwidth of $\theta = \SI{60}{\degree}$.

\begin{figure}
    \centering
    \setlength{\figurewidth}{0.4\columnwidth}
    \setlength{\figureheight}{0.8\figurewidth}
    \subfloat[Padova.\label{fig:oneWebBeamPadova}]{

\definecolor{mycolor1}{RGB}{215,48,39}%
\definecolor{mycolor2}{RGB}{252,141,89}%
\definecolor{mycolor3}{RGB}{254,224,139}%
\definecolor{mycolor4}{RGB}{217,239,139}%
\definecolor{mycolor5}{RGB}{145,207,96}%
\definecolor{mycolor6}{RGB}{26,152,80}%

\begin{tikzpicture}

\begin{axis}[%
width=0.951\figurewidth,
height=\figureheight,
scale only axis,
xmin=0,
xmax=30,
xlabel={N, \# of SV in view},
ymin=0,
ymax=1,
ytick distance=0.1,
ylabel={$P(n>N)$},
xmajorgrids,
ymajorgrids,
enlargelimits=false,title style={font=\scriptsize},xlabel style={font=\scriptsize},ylabel style={font=\scriptsize},ticklabel style={font=\scriptsize},
legend style={at={(0.97,0.97)}, anchor=north east, font=\scriptsize, legend cell align=center, align=left, draw=white!15!black}
]
\addplot [color=mycolor1, line width=1.5pt, mark=o, mark options={solid, mycolor1}]
  table[row sep=crcr]{%
0	1\\
1	0.346981263011798\\
2	0.0277585010409442\\
3	0.00346981263011781\\
4	0\\
};
\addlegendentry{$\varphi=\SI{30}{\degree}$}

\addplot [color=mycolor2, line width=1.5pt, mark=o, mark options={solid, mycolor2}]
  table[row sep=crcr]{%
0	1\\
1	0.612768910478835\\
2	0.0832755031228318\\
3	0.00971547536433004\\
4	0\\
};
\addlegendentry{$\varphi=\SI{40}{\degree}$}
\addplot [color=mycolor3, line width=1.5pt, mark=o, mark options={solid, mycolor3}]
  table[row sep=crcr]{%
0	1\\
1	0.866065232477446\\
2	0.267869535045108\\
3	0.0346981263011799\\
4	0.00485773768216546\\
5	0\\
};
\addlegendentry{$\varphi=\SI{50}{\degree}$}
\addplot [color=mycolor4, line width=1.5pt, mark=o, mark options={solid, mycolor4}]
  table[row sep=crcr]{%
0	1\\
1	0.994448299791811\\
2	0.676613462873005\\
3	0.147814018043026\\
4	0.0215128383067311\\
5	0.00277585010409442\\
6	0\\
};
\addlegendentry{$\varphi=\SI{60}{\degree}$}
\addplot [color=mycolor5, line width=1.5pt, mark=o, mark options={solid, mycolor5}]
  table[row sep=crcr]{%
1	1\\
2	0.981263011797363\\
3	0.566967383761277\\
4	0.201943095072866\\
5	0.0464954892435809\\
6	0.00416377515614119\\
7	0\\
};
\addlegendentry{$\varphi=\SI{70}{\degree}$}
\addplot [color=mycolor6, line width=1.5pt, mark=o, mark options={solid, mycolor6}]
  table[row sep=crcr]{%
1	1\\
2	0.99791811242193\\
3	0.9500346981263\\
4	0.759888965995836\\
5	0.409437890353921\\
6	0.141568355308813\\
7	0.0256766134628723\\
8	0.00208188757807015\\
9	0\\
};
\addlegendentry{$\varphi=\SI{80.92}{\degree}$}
\end{axis}

\end{tikzpicture}%
}
    \subfloat[Svalbard.\label{fig:oneWebBeamSvalbard}]{

\definecolor{mycolor1}{RGB}{215,48,39}%
\definecolor{mycolor2}{RGB}{252,141,89}%
\definecolor{mycolor3}{RGB}{254,224,139}%
\definecolor{mycolor4}{RGB}{217,239,139}%
\definecolor{mycolor5}{RGB}{145,207,96}%
\definecolor{mycolor6}{RGB}{26,152,80}%

\begin{tikzpicture}

\begin{axis}[%
width=0.951\figurewidth,
height=\figureheight,
scale only axis,
xmin=0,
xmax=30,
xlabel={N, \# of SV in view},
ymin=0,
ymax=1,
ytick distance=0.1,
ylabel={$P(n>N)$},
axis background/.style={fill=white},
xmajorgrids,
ymajorgrids,
enlargelimits=false,title style={font=\scriptsize},xlabel style={font=\scriptsize},ylabel style={font=\scriptsize},ticklabel style={font=\scriptsize},
legend style={at={(0.97,0.97)}, anchor=north east, font=\scriptsize, legend cell align=center, align=left, draw=white!15!black}
]
\addplot [color=mycolor1, line width=1.5pt, mark=o, mark options={solid, mycolor1}]
  table[row sep=crcr]{%
0	1\\
1	0.851492019430951\\
2	0.38792505204719\\
3	0.0693962526023597\\
4	0.0117973629424011\\
5	0.000693962526023384\\
6	0\\
};
\addlegendentry{$\varphi=\SI{30}{\degree}$}
\addplot [color=mycolor2, line width=1.5pt, mark=o, mark options={solid, mycolor2}]
  table[row sep=crcr]{%
0	1\\
1	0.989590562109646\\
2	0.882720333102013\\
3	0.429562803608604\\
4	0.148507980569049\\
5	0.0409437890353921\\
6	0.00555170020818885\\
7	0.000693962526023384\\
8	0\\
};
\addlegendentry{$\varphi=\SI{40}{\degree}$}
\addplot [color=mycolor3, line width=1.5pt, mark=o, mark options={solid, mycolor3}]
  table[row sep=crcr]{%
1	1\\
2	0.986814712005552\\
3	0.918112421929216\\
4	0.678695350451076\\
5	0.377515614156836\\
6	0.163081193615545\\
7	0.0645385149201942\\
8	0.0222068008327554\\
9	0.00555170020818885\\
10	0.000693962526023384\\
11	0\\
};
\addlegendentry{$\varphi=\SI{50}{\degree}$}
\addplot [color=mycolor4, line width=1.5pt, mark=o, mark options={solid, mycolor4}]
  table[row sep=crcr]{%
1	1\\
2	0.999306037473975\\
3	0.996530187369881\\
4	0.96946564885496\\
5	0.89590562109646\\
6	0.751561415683554\\
7	0.527411519777932\\
8	0.290076335877863\\
9	0.16030534351145\\
10	0.0922970159611367\\
11	0.0458015267175576\\
12	0.0145732130464964\\
13	0.00832755031228416\\
14	0.00277585010409354\\
15	0.00069396252602516\\
16	0\\
};
\addlegendentry{$\varphi=\SI{60}{\degree}$}
\addplot [color=mycolor5, line width=1.5pt, mark=o, mark options={solid, mycolor5}]
  table[row sep=crcr]{%
3	1\\
4	0.999306037473975\\
5	0.997918112421928\\
6	0.989590562109647\\
7	0.954198473282442\\
8	0.893823733518389\\
9	0.807078417765442\\
10	0.672449687716863\\
11	0.485079805690493\\
12	0.333795975017349\\
13	0.218598195697432\\
14	0.154753643303263\\
15	0.0909090909090899\\
16	0.0464954892435792\\
17	0.015267175572518\\
18	0.00624566273421223\\
20	0.00069396252602516\\
21	0\\
};
\addlegendentry{$\varphi=\SI{70}{\degree}$}
\addplot [color=mycolor6, line width=1.5pt, mark=o, mark options={solid, mycolor6}]
  table[row sep=crcr]{%
7	1\\
8	0.999306037473975\\
9	0.997224149895906\\
10	0.991672449687716\\
11	0.968771686328939\\
12	0.93060374739764\\
13	0.875780707841777\\
14	0.803608605135324\\
15	0.74323386537127\\
16	0.675225537820957\\
17	0.596113809854266\\
18	0.510062456627342\\
19	0.433032616238723\\
20	0.373351839000694\\
21	0.318528799444831\\
22	0.265093684941014\\
23	0.208882720333101\\
24	0.151283830673144\\
25	0.0978487161693273\\
26	0.0555170020818885\\
27	0.0270645385149209\\
28	0.0117973629424029\\
29	0.00277585010409354\\
31	0.00069396252602516\\
};
\addlegendentry{$\varphi=\SI{80.92}{\degree}$}
\end{axis}
\end{tikzpicture}%
}\\
    \subfloat[La Reunion.\label{fig:oneWebBeamReunion}]{\definecolor{mycolor1}{RGB}{215,48,39}%
\definecolor{mycolor2}{RGB}{252,141,89}%
\definecolor{mycolor3}{RGB}{254,224,139}%
\definecolor{mycolor4}{RGB}{217,239,139}%
\definecolor{mycolor5}{RGB}{145,207,96}%
\definecolor{mycolor6}{RGB}{26,152,80}%

\begin{tikzpicture}

\begin{axis}[%
width=0.951\figurewidth,
height=\figureheight,
scale only axis,
xmin=0,
ytick distance=0.1,
xmax=30,
xlabel={N, \# of SV in view},
ymin=0,
ymax=1,
ylabel={$P(n>N)$},
xmajorgrids,
ymajorgrids,
enlargelimits=false,title style={font=\scriptsize},xlabel style={font=\scriptsize},ylabel style={font=\scriptsize},ticklabel style={font=\scriptsize},
legend style={at={(0.97,0.97)}, anchor=north east, font=\scriptsize, legend cell align=center, align=left, draw=white!15!black}
]
\addplot [color=mycolor1, line width=1.5pt, mark=o, mark options={solid, mycolor1}]
  table[row sep=crcr]{%
0	1\\
1	0.252602359472589\\
2	0.0145732130464955\\
3	0\\
};
\addlegendentry{$\varphi=\SI{30}{\degree}$}

\addplot [color=mycolor2, line width=1.5pt, mark=o, mark options={solid, mycolor2}]
  table[row sep=crcr]{%
0	1\\
1	0.458709229701596\\
2	0.0499653018736987\\
3	0.000693962526023384\\
4	0\\
};
\addlegendentry{$\varphi=\SI{40}{\degree}$}
\addplot [color=mycolor3, line width=1.5pt, mark=o, mark options={solid, mycolor3}]
  table[row sep=crcr]{%
0	1\\
1	0.666204024982651\\
2	0.181818181818182\\
3	0.0166551006245665\\
4	0\\
};
\addlegendentry{$\varphi=\SI{50}{\degree}$}
\addplot [color=mycolor4, line width=1.5pt, mark=o, mark options={solid, mycolor4}]
  table[row sep=crcr]{%
0	1\\
1	0.864677307425399\\
2	0.438584316446912\\
3	0.0589868147120054\\
4	0.00277585010409442\\
5	0\\
};
\addlegendentry{$\varphi=\SI{60}{\degree}$}
\addplot [color=mycolor5, line width=1.5pt, mark=o, mark options={solid, mycolor5}]
  table[row sep=crcr]{%
1	1\\
2	0.850798056904927\\
3	0.206800832755031\\
4	0.031228313671062\\
5	0.00346981263011781\\
6	0\\
};
\addlegendentry{$\varphi=\SI{70}{\degree}$}
\addplot [color=mycolor6, line width=1.5pt, mark=o, mark options={solid, mycolor6}]
  table[row sep=crcr]{%
1	1\\
2	0.993754337265788\\
3	0.763358778625955\\
4	0.334489937543372\\
5	0.0922970159611385\\
6	0.0229007633587788\\
7	0.0034698126301187\\
8	0\\
};
\addlegendentry{$\varphi=\SI{80.92}{\degree}$}
\end{axis}

\end{tikzpicture}%
}
    \subfloat[ESTEC.\label{fig:oneWebBeamESTEC}]{

\definecolor{mycolor1}{RGB}{215,48,39}%
\definecolor{mycolor2}{RGB}{252,141,89}%
\definecolor{mycolor3}{RGB}{254,224,139}%
\definecolor{mycolor4}{RGB}{217,239,139}%
\definecolor{mycolor5}{RGB}{145,207,96}%
\definecolor{mycolor6}{RGB}{26,152,80}%

\begin{tikzpicture}

\begin{axis}[%
width=0.951\figurewidth,
height=\figureheight,
at={(0\figurewidth,0\figureheight)},
scale only axis,
xmin=0,
xmax=30,
xlabel={N, \# of SV in view},
ytick distance=0.1,
ymin=0,
ymax=1,
ylabel={$P(n>N)$},
xmajorgrids,
ymajorgrids,
enlargelimits=false,title style={font=\scriptsize},xlabel style={font=\scriptsize},ylabel style={font=\scriptsize},ticklabel style={font=\scriptsize},
legend style={at={(0.97,0.97)}, anchor=north east, font=\scriptsize, legend cell align=center, align=left, draw=white!15!black}
]
\addplot [color=mycolor1, line width=1.5pt, mark=o, mark options={solid, mycolor1}]
  table[row sep=crcr]{%
0	1\\
1	0.252602359472589\\
2	0.0145732130464955\\
3	0\\
};
\addlegendentry{$\varphi=\SI{30}{\degree}$}
\addplot [color=mycolor2, line width=1.5pt, mark=o, mark options={solid, mycolor2}]
  table[row sep=crcr]{%
0	1\\
1	0.458709229701596\\
2	0.0499653018736987\\
3	0.000693962526023384\\
4	0\\
};
\addlegendentry{$\varphi=\SI{40}{\degree}$}
\addplot [color=mycolor3, line width=1.5pt, mark=o, mark options={solid, mycolor3}]
  table[row sep=crcr]{%
0	1\\
1	0.666204024982651\\
2	0.181818181818182\\
3	0.0166551006245665\\
4	0\\
};
\addlegendentry{$\varphi=\SI{50}{\degree}$}
\addplot [color=mycolor4, line width=1.5pt, mark=o, mark options={solid, mycolor4}]
  table[row sep=crcr]{%
0	1\\
1	0.864677307425399\\
2	0.438584316446912\\
3	0.0589868147120054\\
4	0.00277585010409442\\
5	0\\
};
\addlegendentry{$\varphi=\SI{60}{\degree}$}
\addplot [color=mycolor5, line width=1.5pt, mark=o, mark options={solid, mycolor5}]
  table[row sep=crcr]{%
1	1\\
2	0.850798056904927\\
3	0.206800832755031\\
4	0.031228313671062\\
5	0.00346981263011781\\
6	0\\
};
\addlegendentry{$\varphi=\SI{70}{\degree}$}
\addplot [color=mycolor6, line width=1.5pt, mark=o, mark options={solid, mycolor6}]
  table[row sep=crcr]{%
1	1\\
2	0.993754337265788\\
3	0.763358778625955\\
4	0.334489937543372\\
5	0.0922970159611385\\
6	0.0229007633587788\\
7	0.0034698126301187\\
8	0\\
};
\addlegendentry{$\varphi=\SI{80.92}{\degree}$}
\end{axis}

\end{tikzpicture}%
}
    \caption{\ac{ccdf} of the number of OneWeb \acp{sv} in view, for different $\varphi$ and $\theta = \SI{40}{\degree}$, with scenario described in Tab. \ref{tab:simParam2}.}
    \label{fig:oneWEbBeam}
\end{figure}
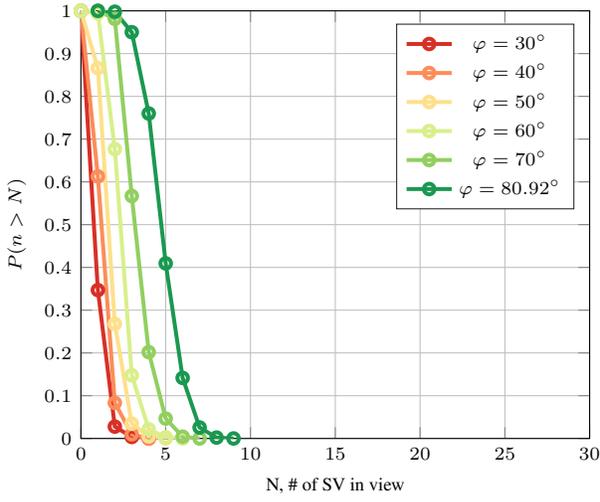
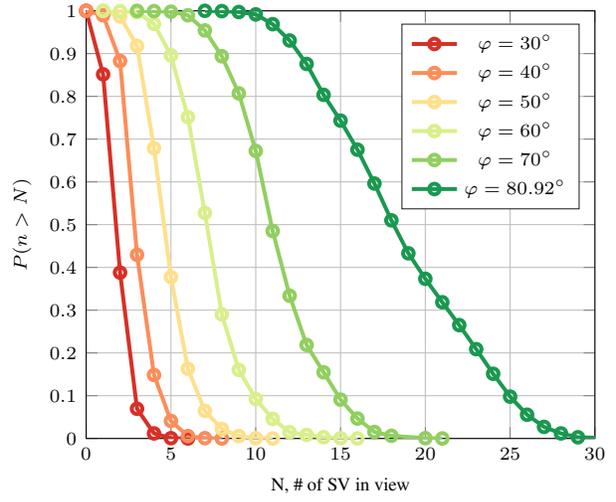
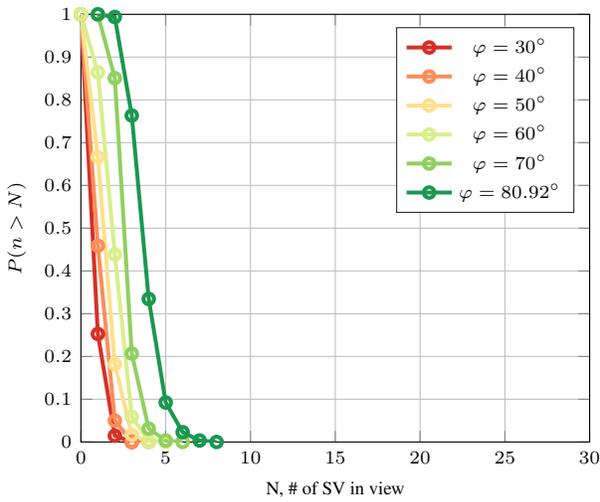
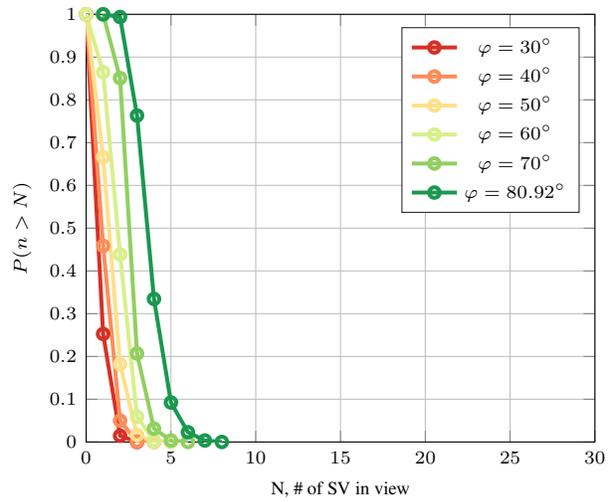

\section{Discussion on Terrestrial SOOP}
Suitable candidates as terrestrial \ac{soop} are i) Bluetooth 4.0 and Bluetooth Low Energy (BLE), ii)  Wi-Fi, and iii) 5G cellular network.

\paragraph{Bluetooth}
Bluetooth 4.0 and BLE have several drawbacks when used for positioning. First, their range is typically limited to around $\SI{10}{\meter}$, restricting their availability and effectiveness in larger spaces. Additionally, the achievable position accuracy is limited since positioning algorithms mainly rely on \ac{rssi} measurements. Lastly, a significant limitation of using Bluetooth 4.0 and BLE as a \ac{soop} is the lack of infrastructure as the positions of Bluetooth beacons are typically not known at the receiver.

\paragraph{Wi-Fi}
Similar drawbacks can be associated with Wi-Fi positioning, which can guarantee good performance in terms of accuracy only in an indoor environment. Wi-Fi networks are typically designed for indoor use, resulting in limited coverage and signal strength in outdoor environments.
In an outdoor scenario, a solution adopted for positioning through Wi-Fi is to use fingerprinting \cite{bisio2019outdoor}, thus creating a database of Wi-Fi signal characteristics at various locations. Each location's unique fingerprint is recorded, capturing signal strength and other parameters from nearby Wi-Fi access points. When positioning is needed, the current WiFi signals are compared to the database to estimate the device's location. This method can provide improved accuracy compared to simple \ac{rssi}-based methods but faces challenges such as environmental variability, signal interference, and the need for extensive initial data collection and maintenance of the fingerprint database.
Finally, effective Wi-Fi positioning requires a dense network of access points with known locations, which may not be available or feasible to deploy in many outdoor areas.

\paragraph{Cellular Networks}
On the other hand, 5G measurements can offer improved reliability, and accuracy in urban scenarios, especially considering the soon-to-be massive deployment of \acp{bs}, and thus anchors for positioning. 
For instance, they may serve as a backup for GNSS when unavailable and enhance overall performance by integrating measurements from both systems \cite{gonzalez2022hybridization}. Differently from both Bluetooth and Wi-Fi, the information about the 5G infrastructure, such as the positions of 5G \acp{bs}, is publicly available. 
Moreover, a new positioning method relying on 5G has been proposed, using the 5G \acp{prs}, defined in TS 38.211 \cite{TS38211}. 
The \acp{prs} are transmitted to the \ac{ue} by multiple \acp{bs}. Next, the \ac{ue} performs measurements for positioning that are then reported back to the network, which calculate the \ac{ue}'s position by using trilateration or triangulation methods. 

The allocation of time-frequency resources for \acp{prs} is adaptable \cite{wei2023prs}, enabling the multiplexing of different downlink PRS signals from multiple \acp{bs} across different subcarriers in a comb-like pattern. Moreover, the comb structure of \ac{prs} manages the interference caused by \ac{prs} signals transmitted by multiple \acp{bs}.
Thus, depending on the application, the requirements may vary and the 5G flexibility will help optimize the resources.
Moreover, apart from \ac{prs}, other 5G synchronization signals (SS), such as the primary SS (PSS), the secondary SS (SSS), and the demodulation reference signals (DMRS), can be used for positioning \cite{tedeschini2023feasibility}, achieving good performance. 

To summarize, the qualitative comparison between all the terrestrial \ac{soop} is provided in Table \ref{tab:terrestrial}.

\begin{table}
\centering
\caption{Qualitative comparison of terrestrial \ac{soop}.}
\begin{tabular}{|l|p{3.5cm}|p{3.5cm}|p{3.5cm}|}
\hline
      & \textbf{Accuracy}                 & \textbf{Availability   }          & \textbf{Supporting Data }        \\ \hline
\textbf{BT}    & Only based on \ac{rssi}.  &  BLE and BT 4.0 have limited range. &  No infrastructure. The positions of BT beacons are not known.\\ \hline
\textbf{5G }   &  High PVT accuracy with PRS. &  Extensive coverage in urban environments, good coverage in rural areas. & Data on BSs positions are available and collected in a dataset.\\ \hline
\textbf{Wi-Fi} & Based on \ac{rssi} or fingerprinting. &  Mainly Indoor. Outdoor-only fingerprinting.&  In fingerprinting WiFi, the AP position is known only within an area of 100m. \\ \hline
\end{tabular}\label{tab:terrestrial}
\end{table}

\section{Conclusions}
In this paper, we have investigated the potential of non-terrestrial and terrestrial \ac{soop} for navigation applications. Our analysis of non-terrestrial \ac{soop} has employed \ac{mcrlb} to establish a relationship between \ac{soop} characteristics and the accuracy of ranging information. We have computed a lower bound on the ranging accuracy using \ac{mcrlb}, considering the estimation of propagation delay, frequency offset, phase offset, and \ac{aoa} across various non-terrestrial \ac{soop} candidates. 
Additionally, we assessed \ac{gdop} and the availability of \ac{leo} SOOP. Our validation process involved comparing \ac{mcrlb} predictions with actual ranging measurements obtained in realistic simulated scenarios. 
Furthermore, we qualitatively evaluated various terrestrial \ac{soop}, considering aspects such as signal availability, achievable accuracy, and infrastructure requirements.

Based on the comprehensive investigation into non-terrestrial and terrestrial \ac{soop} for navigation applications, this study concludes that Starlink and the 5G cellular network emerge as the optimal choices. The Starlink \ac{leo} constellation demonstrates significant advantages in accuracy of ranging information.
The terrestrial \ac{soop} evaluation highlights the suitability of the 5G cellular network, emphasizing its widespread signal availability, achievable accuracy, and existing infrastructure support. These qualities position 5G as a strong contender for terrestrial navigation applications.

\section*{Acknowledgements}

This work was partially funded by the European Space Agency under contract n. 4000143575/24/NL/WC/kg: “Navigation Using Machine lEaRning applied to Signals of Opportunity (NUMEROSO)”. This work was also partially supported by the European Union under the Italian National Recovery and Resilience Plan (PNRR) of NextGenerationEU, partnership on “Telecommunications of the Future” (PE0000001 - program ``RESTART'').

\bibliographystyle{apalike}
\bibliography{biblio}

\begin{thebibliography}{}

\bibitem[3GPP, 2020]{TS38211}
3GPP (2020).
\newblock {5G}; {NR}; physical channels and modulations.
\newblock Technical Report TS 38.211 V 16.2.0, 3GPP.

\bibitem[Aguilar et~al., 2019]{Aguilar19Tradespace}
Aguilar, A., Butler, P., Collins, J., Guerster, M., Kristinsson, B., Mckeen, P., Cahoy, K., and Crawley, E. (2019).
\newblock Tradespace exploration of the next generation communication satellites.
\newblock In {\em AIAA Scitech Forum}.

\bibitem[Bisio et~al., 2019]{bisio2019outdoor}
Bisio, I., Garibotto, C., Lavagetto, F., and Sciarrone, A. (2019).
\newblock Outdoor places of interest recognition using {Wi-Fi} fingerprints.
\newblock {\em IEEE Transactions on Vehicular Technology}, 68(5):5076--5086.

\bibitem[D'Andrea et~al., 1994]{dandrea94modified}
D'Andrea, A., Mengali, U., and Reggiannini, R. (1994).
\newblock The modified {C}ramer-{R}ao bound and its application to synchronization problems.
\newblock {\em IEEE Transactions on Communications}, 42(234):1391--1399.

\bibitem[del Portillo~Barrios et~al., 2019]{delportillo2019technical}
del Portillo~Barrios, I., Cameron, B., and Crawley, E. (2019).
\newblock A technical comparison of three low {E}arth orbit satellite constellation systems to provide global broadband.
\newblock {\em Acta Astronautica}, 159.

\bibitem[Dey, 2014]{iridium_link_budget}
Dey, S. (2014).
\newblock An approach to calculate the performance and link budget of {LEO} satellite ({Iridium}) for communication operated at frequency range (1650-1550) {MHz}.
\newblock {\em International Journal of Latest Trends in Engineering and Technology}, 4:96--103.

\bibitem[Dey et~al., 2023]{formula_link_budget}
Dey, S., Professor, A., and Baishali, B. (2023).
\newblock Link budget of {LEO} satellite (sky bridge) for communication operated at {Ku} band frequency range (12-14) {GHz}.
\newblock {\em International Journal of Innovations in Engineering and Technology (IJIET)}, 4:173--179.

\bibitem[{Eutelsat OneWeb}, 2017]{oneWebTISS}
{Eutelsat OneWeb} (2017).
\newblock One{W}eb non-geostationary satellite system attachment {A}: {T}echnical information to supplement schedule {S}.
\newblock \url{https://fcc.report/IBFS/SAT-LOI-20160428-00041/1134939}.
\newblock Last Access: May 2024.

\bibitem[Ferre and Lohan, 2021]{ferre2021comparison}
Ferre, R.~M. and Lohan, E.~S. (2021).
\newblock Comparison of {MEO}, {LEO}, and terrestrial {IoT} configurations in terms of {GDOP} and achievable positioning accuracies.
\newblock {\em IEEE Journal of Radio Frequency Identification}, 5(3):287--299.

\bibitem[Ferre et~al., 2022]{ferre2022isleo}
Ferre, R.~M., Lohan, E.~S., Kuusniemi, H., Praks, J., Kaasalainen, S., Pinell, C., and Elsanhoury, M. (2022).
\newblock Is {LEO}-based positioning with mega-constellations the answer for future equal access localization?
\newblock {\em IEEE Communications Magazine}, 60(6):40--46.

\bibitem[Gonzalez-Garrido et~al., 2022]{gonzalez2022hybridization}
Gonzalez-Garrido, A., Querol, J., and Chatzinotas, S. (2022).
\newblock Hybridization of {GNSS} and {5G} measurements for assured positioning, navigation and timing.
\newblock In {\em Proc. of the 35th International Technical Meeting of the Satellite Division of The Institute of Navigation (ION GNSS+ 2022)}, pages 2377--2384, Denver, CO.

\bibitem[Humphreys et~al., 2023]{Humphreys23Signal}
Humphreys, T.~E., Iannucci, P.~A., Komodromos, Z.~M., and Graff, A.~M. (2023).
\newblock Signal structure of the {S}tarlink {K}u-band downlink.
\newblock {\em IEEE Transactions on Aerospace and Electronic Systems}, 59(5):6016--6030.

\bibitem[ICAO, 2007]{iridium_manual}
ICAO (2007).
\newblock {ICAO} technical manual for {I}ridium aeronautical mobile satellite (route) service.
\newblock \url{https://www.icao.int/safety/acp/Inactive\%20working\%20groups\%20library/ACP-WG-M-Iridium-8/IRD-SWG08-IP05\%20-\%20AMS(R)S\%20Manual\%20Part%20II\%20v4.0.pdf}.
\newblock Last Access: June 2024.

\bibitem[Kay, 1997]{Kay97}
Kay, S.~M. (1997).
\newblock {\em Fundamentals of Statistical Signal Processing: Estimation Theory}.
\newblock Prentice Hall.

\bibitem[Komodromos et~al., 2023]{komodromos2023IONsimulator}
Komodromos, Z.~M., Qin, W., and Humphreys, T.~E. (2023).
\newblock Signal simulator for {Starlink} {Ku-Band} downlink.
\newblock In {\em Institute of Navigation GNSS+ Conference}.

\bibitem[Kozhaya et~al., 2023]{10139969}
Kozhaya, S., Kanj, H., and Kassas, Z.~M. (2023).
\newblock Multi-constellation blind beacon estimation, {D}oppler tracking, and opportunistic positioning with {OneWeb}, {Starlink}, {Iridium NEXT}, and {Orbcomm} {LEO} satellites.
\newblock In {\em Proc. of IEEE/ION Position, Location and Navigation Symposium (PLANS)}, pages 1184--1195.

\bibitem[Mengali and D'Andrea, 1997]{mengali1997synchronization}
Mengali, U. and D'Andrea, A.~N. (1997).
\newblock {\em {Synchronization techniques for digital receivers}}.
\newblock Springer.

\bibitem[ORBCOMM, 2001]{orbcomm_manuale}
ORBCOMM (2001).
\newblock {ORBCOMM} system overview, {A80TD0008} - revision {G}.
\newblock \url{https://ctu.gov.cz/sites/default/files/cs/download/oznamene_typy_rozhrani/orbcomm-rozhrani_02_06_2010.pdf}.
\newblock Last Access: June 2024.

\bibitem[Osoro and Oughton, 2021]{Osoro21Techno}
Osoro, O.~B. and Oughton, E.~J. (2021).
\newblock A techno-economic framework for satellite networks applied to low {E}arth orbit constellations: Assessing {S}tarlink, {OneWeb} and {Kuiper}.
\newblock {\em IEEE Access}, 9:141611--141625.

\bibitem[Reid et~al., 2016]{Reid2016LeveragingCB}
Reid, T. G.~R., Neish, A.~M., Walter, T., and Enge, P.~K. (2016).
\newblock Leveraging commercial broadband {LEO} constellations for navigating.
\newblock In {\em Proc. of the 29th International Technical Meeting of the Satellite Division of The Institute of Navigation (ION GNSS+ 2016)}.

\bibitem[SpaceX, 2016]{starlinkTISS}
SpaceX (2016).
\newblock Space{X} non-geostationary satellite system attachment {A}: {T}echnical information to supplement schedule {S}.
\newblock \url{https://fcc.report/IBFS/SAT-LOA-20161115-00118/1158350}.
\newblock Last Access: May 2024.

\bibitem[Tedeschini et~al., 2023]{tedeschini2023feasibility}
Tedeschini, B.~C., Brambilla, M., Italiano, L., Reggiani, S., Vaccarono, D., Alghisi, M., Benvenuto, L., Realini, A. G.~E., Grec, F., and Nicoli, M. (2023).
\newblock A feasibility study of {5G} positioning with current cellular network deployment.
\newblock {\em Scientific Reports}, 13(15281).

\bibitem[Wei et~al., 2023]{wei2023prs}
Wei, Z., Wang, Y., Ma, L., Yang, S., Feng, Z., Pan, C., Zhang, Q., Wang, Y., Wu, H., and Zhang, P. (2023).
\newblock {5G} {PRS}-based sensing: A sensing reference signal approach for joint sensing and communication system.
\newblock {\em IEEE Transactions on Vehicular Technology}, 72(3):3250--3263.

\bibitem[Xia et~al., 2019]{xia2019beam}
Xia, S., Jiang, Q., Zou, C., and Li, G. (2019).
\newblock Beam coverage comparison of {LEO} satellite systems based on user diversification.
\newblock {\em IEEE Access}, PP:1--1.

\end{thebibliography}

\end{document}